\documentclass[pre,twocolumn,amsmath,amssymb,nofootinbib,floatfix,superscriptaddress]{revtex4}

\usepackage{enumerate}
\usepackage{graphicx,bm}
\usepackage[normalem]{ulem}

\usepackage{graphicx,bm,xcolor}
\usepackage{enumerate}
\usepackage{graphicx,bm}
\usepackage[normalem]{ulem}

\begin{document}

\title{Deconfinement transitions in three-dimensional compact lattice
  Abelian Higgs models with multiple-charge scalar fields}

\author{Claudio Bonati} 
\affiliation{Dipartimento di Fisica dell'Universit\`a di Pisa
        and INFN Sezione di Pisa, Largo Pontecorvo 3, I-56127 Pisa, Italy}

\author{Andrea Pelissetto}
\affiliation{Dipartimento di Fisica dell'Universit\`a di Roma Sapienza
        and INFN Sezione di Roma I, I-00185 Roma, Italy}

\author{Ettore Vicari} 
\affiliation{Dipartimento di Fisica dell'Universit\`a di Pisa,
        Largo Pontecorvo 3, I-56127 Pisa, Italy}

\date{\today}

\begin{abstract}
We investigate the nature of the deconfinement transitions in
three-dimensional lattice Abelian Higgs models, in which a complex
scalar field of integer charge $Q\ge 2$ is minimally coupled with a
compact $U(1)$ gauge field. Their phase diagram presents two phases
separated by a transition line where static charges $q$, with $q<Q$,
deconfine. We argue that these deconfinement transitions belong to the
same universality class as transitions in generic three-dimensional
${\mathbb Z}_Q$ gauge models.  In particular, they are Ising-like for
$Q=2$, of first order for $Q=3$, and belong to the three-dimensional
gauge $XY$ universality class for $Q\ge 4$. This general scenario is
supported by numerical finite-size scaling analyses of the energy
cumulants for $Q=2$, $Q=4$, and $Q=6$.
\end{abstract}

\maketitle

% ========================= BODY =========================

\section{Introduction}
\label{intro}

Three-dimensional (3D) Abelian Higgs (AH) gauge models, in which
charged scalar fields are coupled with an Abelian gauge field, provide
effective models for several emergent collective phenomena in
condensed-matter physics~\cite{Anderson-book,Wen-book,Sachdev-19}. The
phase structure of this class of models has been extensively studied,
see, e.g., Refs.~\cite{HLM-74,FS-79,DH-81,FMS-81,DHMNP-81,BF-81,CC-82,
  BF-83,FM-83,KK-85,KK-86,BN-86,BN-86-b,BN-87,RS-90,MS-90,KKS-94,BFLLW-96,
  HT-96,FH-96,IKK-96,KKLP-98,OT-98, CN-99, HS-00, SSS-02, SM-02,
  KNS-02,MHS-02, SSSNH-02,SSNHS-03,MZ-03,NRR-03, MV-04,SBSVF-04,
  NSSS-04, SSS-04, HW-05, WBJSS-05, CFIS-05, TIM-05, TIM-06, CIS-06,
  KPST-06,Herbut-book, Sandvik-07, WBJS-08, MK-08, JNCW-08, MV-08,
  KMPST-08, CAP-08, KS-08, ODHIM-09, LSK-09, CGTAB-09, CA-10, BDA-10,
  Sandvik-10, GS-10, IMH-12, Kaul-12, KS-12, BMK-13, HBBS-13,
  Bartosch-13, HSOMLWTK-13, CHDKPS-13, PDA-13, BS-13, NCSOS-15,
  NSCOS-15, SP-15, SGS-16,WNMXS-17, FH-17, PV-19-CP, IZMHS-19,
  PV-19-AH3d, SN-19, Sachdev-19, PV-20-largeNCP, SZ-20, PV-20-mfcp,
  BPV-20-hcAH, BPV-21, BPV-21-bgs, BPV-21-ncAH, WB-21, BPV-22-mpf,
  BPV-22, BPV-22-dis, BPV-23b,SZJSM-23,BPV-23d,BPV-24}. It has been
realized that a crucial role is played by topological properties.
Indeed, the phase diagram and the nature of the transitions depend on
the compact or noncompact nature of the gauge
fields~\cite{FS-79,DH-81,KK-85,BN-87, NRR-03, BPV-22}, and, in compact
lattice models, on the charge of the scalar
fields~\cite{FS-79,BF-81,SSNHS-03,BPV-20-hcAH} and on the presence or
absence of topological defects such as
monopoles~\cite{MS-90,MV-04,PV-19-CP,PV-20-mfcp}.

We consider compact lattice AH (CLAH) models defined on 3D cubic
lattices. The partition function reads
\begin{eqnarray}
&&Z = \sum_{\{z,\lambda\}} e^{-\beta H}, \qquad H = J H_z + K
H_\lambda, \label{partfun}\\
&&H_z =  - 2 \sum_{{\bm x}, \mu} {\rm Re}\, 
\bar{z}_{\bm x} \lambda_{{\bm x},\mu}^Q z_{{\bm x}+\hat\mu},
\label{Hzdef}\\
&&H_\lambda = - 2 \sum_{{\bm x},\mu>\nu} {\rm Re}\,
\Pi_{{\bm x},\mu\nu}, 
\label{Hgdef}\\
&&\Pi_{{\bm x},\mu\nu} = \lambda_{{\bm
      x},{\mu}} \,\lambda_{{\bm x}+\hat{\mu},{\nu}}
  \,\bar{\lambda}_{{\bm x}+\hat{\nu},{\mu}} \,\bar{\lambda}_{{\bm
      x},{\nu}},\label{plaquette}
\end{eqnarray}
where $\lambda_{{\bm x},\mu}\in U(1)$ are phases (unit-length complex
variables) associated with each lattice link (starting at site ${\bm
  x}$ along one of the lattice direction, $\mu=1,2,3$), $z_{\bm x}$
are unit-length complex variables associated with each lattice site,
and $Q\in\mathbb{N}$ is the charge of the complex scalar field $z_{\bm
  x}$. Without loss of generality, the inverse temperature
$\beta=1/(k_B T)$ can be absorbed in the definition of the couplings
$J$ and $K$, formally setting $\beta=1$ in the partition function
(\ref{partfun}). The model is invariant under U(1) gauge
transformations, $z_{\bm x} \to U_{\bm x} z_{\bm x}$, $\lambda_{{\bm
    x},\mu} = U_{\bm x} \lambda_{{\bm x},\mu} \bar{U}_{{\bm
    x}+\hat{\mu}}$, where $U_{\bm x}$ is a phase.  The CLAH model can
also be rewritten in terms of the gauge fields only, by exploiting the
gauge invariance. Indeed, by appropriate gauge transformations, we can
set $z_{\bm x} = 1$ everywhere (the so-called unitary gauge),
obtaining the unitary-gauge CLAH Hamiltonian
\begin{equation}
H_{\rm ug} = - 2 J \sum_{{\bm x}\mu} \hbox{Re}\ \lambda_{{\bm
    x},\mu}^Q + K H_\lambda,
\label{Hunitary}
\end{equation}
which presents a residual invariance under local ${\mathbb Z}_Q$ gauge
transformations, i.e., under transformations $\lambda_{{\bm x},\mu} =
V_{\bm x} \lambda_{{\bm x},\mu} \bar{V}_{{\bm x}+\hat{\mu}}$ with
$V_{\bm x}\in {\mathbb Z}_Q$, i.e., such that $V_{\bm x}^Q=1$.

\begin{figure}[tbp]
  \includegraphics*[width=0.90\columnwidth]{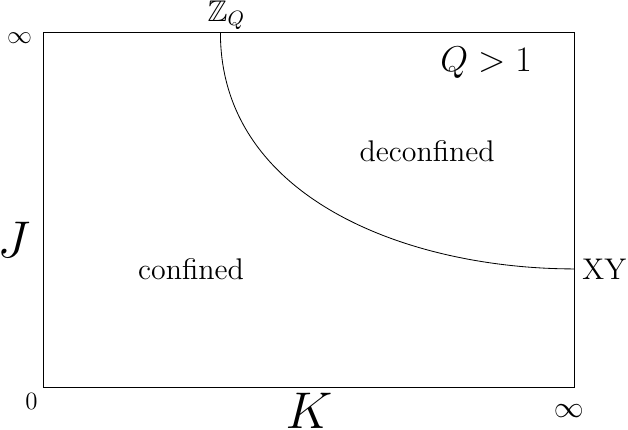}
  \caption{Sketch of the $J$-$K$ phase diagram of the 3D one-component
    CLAH model, in which a compact $U(1)$ gauge field is coupled with
    a (unit-length) complex scalar field with charge $Q\ge 2$.  A
    confined phase (for small $J$ or small $K$) and a deconfined phase
    (for large $J$ and $K$) are present, separated by a deconfinement
    transition line.  For $K\to\infty$ and $J\to\infty$ the model
    reduces to the $XY$ vector model and to the lattice ${\mathbb
      Z}_Q$ gauge model with Wilson action, respectively.  See
    Sec.~\ref{phasediagram} for more details.}
\label{phadia}
\end{figure}

The phase diagrams of the CLAH models with $Q=1$ and $Q>1$ differ
significantly~\cite{FS-79,SSNHS-03}.  While the $Q=1$ model has only
one thermodynamic phase, for $Q\ge 2$ two different phases occur. They
are divided by a transition line where the charged degrees of freedom
with $q<Q$ deconfine.  A sketch of the phase diagram for $Q\ge 2$ is
shown in Fig.~\ref{phadia}.  The deconfinement transition line is
expected to connect the transition points of the models obtained for
$J\to\infty$ and $K\to\infty$, respectively, i.e., the ${\mathbb Z}_Q$
gauge model and the standard $XY$ model~\cite{FS-79,SSNHS-03}.

In the present paper we investigate the nature of the deconfinement
transition line, and, in particular, its universal critical properties
when the transitions are continuous.  Numerical results for the
one-component CLAH model have been reported in
Refs.~\cite{BF-81,SSSNH-02,SSNHS-03} for several values of $Q\ge 2$.
On the basis of analyses of Monte Carlo (MC) data,
Ref.~\cite{SSNHS-03} claimed that the critical behavior along the
deconfinement transition line is not universal, but it is controlled
by a line of fixed points.  So, critical exponents change continuously
along the transition line, varying from those of the ${\mathbb Z}_Q$
gauge transition at $J=\infty$ to those of the $XY$ transition at
$K=\infty$.

This type of behavior, if confirmed, would be quite unusual for 3D
transitions.  Indeed, while there are several examples of fixed-point
lines in two dimensions---exact results have been obtained for the
eight-vertex model~\cite{Baxter-book}, the Ashkin-Teller
model~\cite{AT-43,KW-71,MM-23}, and the low-temperature spin-wave
phase of the $XY$
model~~\cite{KT-73,Berezinskii-70,Kosterlitz-74,JKKN-77}---there is no
robust evidence (i.e., exact results or compelling physical arguments)
that some transition lines in 3D models are associated with lines of
fixed points.  Therefore, it is important to review and improve the
previous numerical results, given that the existence of a line of
fixed points would represent a new phenomenon in three dimensions.

In this paper we return to this issue, performing a detailed
finite-size scaling (FSS) analysis of MC data for $Q=2$, $Q=4$, and
$Q=6$. We anticipate that the results of our analyses are consistent
with the more conventional 3D scenario: The universality class of the
deconfinement transitions does not change along the line. Their
critical behavior is the same as in generic lattice ${\mathbb Z}_{Q}$
gauge models, so the transitions belong to the Ising and $XY$
universality classes for $Q=2$ and $Q\ge 4$, respectively.  This is
the same critical behavior observed in ${\mathbb Z}_{Q}$ gauge models
obtained in the $J\to\infty$ limit with a notable exception, the case
$Q=4$.  This is related to the fact that the ${\mathbb Z}_{4}$ gauge
model obtained for $J\to \infty$ is not a generic ${\mathbb Z}_{4}$
gauge model, but a particular one that is equivalent to two decoupled
Ising gauge models. Instead, along the deconfinement line, the
critical behavior of the $Q=4$ model is expected to be that of a
generic ${\mathbb Z}_{4}$ gauge model, which undergoes an $XY$
transition, as for $Q>4$.

We stress that, since numerical analyses based on MC simulations only
consider a limited range of lattice sizes or correlation lengths (the
same is true for experiments), the control of the nonuniversal scaling
corrections to the asymptotic critical behavior is crucial to check
universality.  Without a proper handling of these corrections,
inaccurate estimates of critical universal parameters are obtained,
leading to apparent violations of universality.  In particular, we
point out the existence of significant crossover effects for large
values of $K$, arising from the unstable $XY$ vector critical behavior
present for $K=\infty$. They may hide the universal behavior when
analyzing numerical data for relatively small lattice sizes.

The paper is organized as follows. In Sec.~\ref{phasediagram} we
discuss the phase diagram of CLAH models with $Q\ge 2$, and put
forward a general scenario for the deconfinement transitions.  In
Sec.~\ref{fssenecum} we outline the main features of our FSS analyses
of the energy cumulants. In Sec.~\ref{numana} we report our FSS
analyses of the MC data for $Q=2$, $Q=4$, and $Q=6$.  In
Sec.~\ref{conclu} we summarize and draw our conclusions. We also
compare the results for the one-component CLAH models with those for
other lattice AH models.  We also add a number of appendices,
reporting useful results that characterize the universality class of
${\mathbb Z}_Q$ gauge models and of gauge $XY$ models.

\section{Phase diagrams of multicharge CLAH models}
\label{phasediagram}

In this section we discuss the phase diagram of the CLAH model defined
in Eq.~(\ref{partfun}) for $Q\ge 2$, as a function of the Hamiltonian
parameters $J$ and $K$ (as mentioned we set $\beta=1$), see
Fig.~\ref{phadia}.  We assume $K \ge 0$, so the parameter $K$ plays
the role of inverse square gauge coupling, i.e., $K \sim
g^{-2}$. Moreover, since the phase diagram is symmetric for $J\to -J$,
as can be seen by using the field redefinition $z_{\bm x}\to
(-1)^{x_1+x_2+x_3}z_{\bm x}$, we assume $J\ge 0$.

For $Q\ge 2$, two different phases are present, which are
distinguished by the confinement properties of the charged excitations
with $q<Q$. Indeed, the transition line separating these two phases is
related to the deconfinement of the charged degrees of freedom with
$q<Q$ and, in particular, to the behavior of the unit-charge Wilson
loops $W_{\cal C} = \prod_{\ell\in \cal C} \lambda_\ell$, where $\cal
C$ is a closed lattice loop.  For $Q \ge 2$, the Wilson loop for
unit-charge sources obeys the area law in the confined phase and the
perimeter law in the deconfined phase.  For $Q=1$ the area law never
holds, due to the screening of the charged scalar modes, so only the
deconfined phase is present.  Note that the nontrivial dependence on
the charge $Q$ is strictly related to the compact gauge action in
which $\lambda_{{\bm x},\mu}\in U(1)$ is used.  In noncompact
formulations, the gauge field is a real field $A_{{\bm x},\mu}$, and
the charge $Q$ can be eliminated by a redefinition of $A_{{\bm
    x},\mu}$ and $K$.

For $J\to\infty$ the model with $Q\ge 2$ is equivalent to the lattice
${\mathbb Z}_Q$ gauge model with Wilson
action~\cite{Wegner-71,SSNHS-03,BPV-20-hcAH}.  Indeed, if we start
from the unitary-gauge Hamiltonian (\ref{Hunitary}), only fields
satisfying $\lambda_{{\bm x},\mu}^Q = 1$ are allowed for
$J\to\infty$. Therefore, we obtain a gauge theory with link variables
\begin{equation}
  \rho_{{\bm x},\mu}=\exp(2\pi i n_{{\bm x},\mu}/Q),
  \qquad n_{{\bm x},\mu}=1,...,Q,
  \label{rhozq}
  \end{equation}
and Hamiltonian
\begin{eqnarray}
H_{{\mathbb Z}_Q}= - 2 K \sum_{{\bm x},\mu>\nu} {\rm Re}
\,(\rho_{{\bm x},{\mu}} \,\rho_{{\bm x}+\hat{\mu},{\nu}}
\,\bar{\rho}_{{\bm x}+\hat{\nu},{\mu}} \,\bar{\rho}_{{\bm x},{\nu}}).
  \label{HZQ}
\end{eqnarray}
The $\mathbb{Z}_Q$ gauge model (\ref{HZQ}) has a topological
transition for a finite $K=K_{{\mathbb
    Z}_Q}$~\cite{Bhanot:1980pc,Borisenko:2013xna}.

The $\mathbb{Z}_Q$ gauge model is dual to a specific $Q$-state clock
model~\cite{Savit-80}, with $\mathbb{Z}_Q$ spin variables $\exp(2\pi i
n_{\bm x}/Q)$ (where $n_{\bm x}=1,...,Q$) associated with the sites,
and a Hamiltonian which is symmetric under global ${\mathbb Z}_Q$
transformations. The 2-state clock model is equivalent to the Ising
model, while the 3-state clock model is equivalent to the 3-state
Potts model, which undergoes a first-order transition. The nature of
the transitions for $Q\ge 4$ can be understood using the
Landau-Ginzburg-Wilson (LGW) approach. The LGW Hamiltonian for a
$\mathbb{Z}_Q$-symmetric spin system can be written
as~\cite{BPV-22-dis}
\begin{eqnarray}
    {\cal H}_{\mathbb{Z}_Q} = |\partial_\mu \Phi|^2 + r\,
    |\Phi|^2 + u \,|\Phi|^4 + v \,(\Phi^Q + \bar\Phi^Q),
\label{lgwzq}
\end{eqnarray}
where $\Phi({\bm x})$ is a complex field.  Note that ${\cal
  H}_{\mathbb{Z}_Q}$ is also invariant under the transformation
$\Phi\to\bar\Phi$, which, however, does not play any role at
criticality.  The $Q$-dependent potential has dimension $Q$ and is
therefore irrelevant for $Q > 4$. We can thus set $v=0$, obtaining the
standard $\Phi^4$ theory for a complex field. We thus predict an $XY$
critical behavior and an effective enlargement of the symmetry at the
transition.

For $Q=4$, the $Q$-dependent potential has dimension four and
represents a cubic anisotropy.  The renormalization-group (RG) flow of
model (\ref{lgwzq}) with $Q=4$ has been studied by using different
approaches, see, e.g., Refs.~\cite{Aharony-76, PV-02, CPV-00,
  Hasenbusch-23}. It turns out that the stable fixed point is again
the $XY$ fixed point with $v=0$.  The anisotropic interaction with
coefficient $v$ gives only rise to scaling corrections.  However, due
to the small absolute value of the corresponding RG
dimension~\cite{CPV-00,CPV-03,HV-11} $y_v=-0.108(6)$, these scaling
corrections are expected to decay slowly. Beside the stable fixed
point along the line $v=0$, the LGW model admits also an unstable
nontrivial fixed point along the line $w=u-6v=0$, where the LGW
Hamiltonian can be written as the sum of two identical LGW
Hamiltonians with a real scalar field, so that this fixed point
corresponds to an Ising critical behavior. The parameter $w$
parametrizes a relevant perturbation of the Ising fixed point with RG
dimension $y_w=d - 2/\nu_I=0.17475(2)$, where~\cite{KPSV-16}
$\nu_I=0.629971(4)$ is the Ising critical exponent.  The LGW analysis
therefore predicts that generic $\mathbb{Z}_4$ invariant clock models
undergo $XY$ transitions.  However, the standard $\mathbb{Z}_4$ clock
model can be exactly rewritten as a sum of two Ising
models~\cite{HS-03}, and thus it exactly corresponds to the LGW theory
with $w=0$, hence it undergoes an Ising transition.
  
We can use duality and the above results for $Q$-state clock models to
predict the nature of the transitions in ${\mathbb Z}_Q$ gauge models,
corresponding to the $J\to\infty$ limit of the CLAH models.  For $Q=2$
we have an Ising transition for $J=\infty$ and $K_{{\mathbb
    Z}_2}=0.380706646(6)$~\cite{Wegner-71,BPV-20-hcAH,FXL-18}, while
for $Q=3$, the transition should be of first order.  The ${\mathbb
  Z}_4$ gauge model is dual to the standard ${\mathbb Z}_4$ clock
model, hence it should undergo an Ising transition.  This result also
follows from the exact relation
\begin{equation}
  Z_{{\mathbb Z}_4}(K) = Z_{{\mathbb Z}_2}(K/2)^2,\quad
  Z_{{\mathbb
  Z}_Q}(K)=\sum_{\{\rho\}}e^{-H_{{\mathbb Z}_Q}(K)},
\label{zkqrel}
\end{equation}
between the partition functions of the ${\mathbb Z}_4$ and ${\mathbb
  Z}_2$ gauge models defined by the Hamiltonian (\ref{HZQ}), which is
demonstrated in App.~\ref{z4z22}. This relation also implies that the
transition point of the ${\mathbb Z}_4$ gauge model is located at
$K_{{\mathbb Z}_4}=2K_{{\mathbb Z}_2} = 0.761413292(12)$.  For $Q>4$
the transitions of the lattice ${\mathbb Z}_Q$ gauge models belong to
the 3D $XY$ universality class~\cite{HS-03,BPV-20-hcAH}. For $Q=6$ the
transition occurs at $K_{{\mathbb Z}_6}=
1.50342(4)$~\cite{Borisenko:2013xna}.  Moreover, for large $Q$ we
have~\cite{BPV-22}
\begin{equation}
K_{{\mathbb Z}_Q} \approx \frac{K_{IXY}}{2} Q^2,\qquad
K_{IXY} =0.076051(2),
\label{KZlQ}
\end{equation}
where $K_{IXY}$ is the critical point of the inverted $XY$ model that
represents the $J\to\infty$ limit of the noncompact lattice AH model.

The previous discussion allows us to conclude that the limiting
${\mathbb Z}_Q$ gauge model with Hamiltonian (\ref{HZQ}) undergoes an
Ising transition for $Q=2$ and 4, and an $XY$ transition for $Q\ge
5$. It is however important to stress that this identification is done
using the duality between the gauge and the spin system, and that
duality only maps the free energy and the related thermal
observables. Strictly speaking, the Ising and $XY$ fixed points that
control the critical behavior of ${\mathbb Z}_Q$ gauge models differ
from the Ising and $XY$ fixed points that are relevant for the
corresponding ${\mathbb Z}_Q$ clock models.  Indeed, they represent
fixed points obtained by performing RG transformations on two
different classes of Hamiltonians---those with local and global
${\mathbb Z}_Q$ symmetry, respectively.  We have used the same name
because of the existence of the duality transformation, which,
however, only relates the thermal sectors of the two models. There are
both relevant and irrelevant operators that are unrelated: for
instance, the magnetic sector in the spin universality class has no
counterpart in the gauge universality class. In the following, to
avoid confusion, we distinguish the gauge $XY$ from the spin $XY$
universality class by writing $XY_G$ in the gauge case.  No suffix
will be used for the Ising case, as no confusion can arise: in this
work we always refer to the Ising-gauge universality class.

For $K\to\infty$, the plaquette term becomes trivial, i.e., $\Pi_{{\bm
    x},\mu\nu}=1$ everywhere. Therefore, in the thermodynamic limit,
we can set $\lambda_{{\bm x},\mu} = 1$ modulo gauge transformations
(in a finite volume there are some subtleties that are discussed in
App.~\ref{App.B}).  The scalar Hamiltonian becomes
\begin{equation}
  H_{XY}(J) =  - 2 J \sum_{{\bm x}, \mu} 
  {\rm Re}\,\bar{z}_{\bm x} z_{{\bm x}+\hat\mu}, \label{HXY}
\end{equation}
for any $Q$. The model undergoes therefore an $XY$ vector transition
at~\cite{Hasenbusch-19,CHPV-06,DBN-05} $J_{XY}=0.22708234(9)$,
for any $Q$. 

The $XY$ spin fixed point that controls the critical behavior in the
model (\ref{HXY}) is unstable with respect to the addition of Abelian
gauge variables~\cite{BPV-22}.  This can be shown by computing the RG
dimension of the gauge operator that drives the system out of the
fixed point in the AH field theory~\cite{BPV-21-ncAH}.  To all orders
of perturbation theory in the $\varepsilon=4-d$ expansion, we
obtain~\cite{BPV-21-ncAH} $y_{\alpha}=4-d$ ($y_{\alpha}= 1$ in three
dimensions) for the crossover exponent associated with the squared
gauge coupling $\alpha\equiv g^2\sim 1/K$.  The RG dimension
$y_\alpha$ corresponds to the energy dimension of the gauge coupling
$\alpha$.  Therefore, in the large-$K$ limit, the gauge field gives
rise to an intrinsic crossover scale $\xi_\alpha \sim K$. If the
correlation length $\xi$ or the size of the system $L$ satisfies $\xi
\lesssim \xi_\alpha$ or $L \lesssim \xi_\alpha$, significant crossover
effects with an apparent $XY$ behavior can be observed.

We now turn to discuss the deconfinement transitions for finite $J$
and $K$.  A natural working hypothesis is that all transitions along
the line running from $[J=\infty,K_{{\mathbb Z}_Q}]$ to
$[J_{XY},K=\infty]$ have the same universal features---as we have
already mentioned, a line of fixed points seems unlikely in 3D.  Given
that the CLAH model is equivalent to a purely gauge model with
${\mathbb Z}_Q$ gauge symmetry in the unitary gauge, see
Eq.~(\ref{Hunitary}), the most natural hypothesis is that the
finite-$J$ deconfinement transitions have the same universal features
as the topological transition in the ${\mathbb Z}_Q$ gauge model
obtained for $J\to \infty$.  In this scenario we would expect Ising
deconfinement transitions for $Q=2$, first-order transitions for
$Q=3$, since the $J\to\infty$ transition is of first order, and $XY_G$
transitions for any $Q>4$.

We do not expect the above scenario to work for $Q=4$.  Indeed, as we
have discussed above, the $\mathbb{Z}_4$ gauge model (\ref{HZQ}) is
not a generic $\mathbb{Z}_4$ gauge model: because of the identity
(\ref{zkqrel}), its partition function is equivalent to that of two
decoupled Ising gauge models.  The addition of the scalar field spoils
this factorization [this is more easily seen using the unitary-gauge
  Hamiltonian (\ref{Hunitary})]: Barring unexpected cancellations, the
finite-$J$ scalar Hamiltonian should represent a relevant perturbation
of the $J\to\infty$ Ising fixed point.  Therefore, the finite-$J$
transitions should be controlled by the stable $XY_G$ fixed point also
for $Q=4$. However, for large $J$ we expect significant crossover
effects due to the nearby Ising gauge transition, given the small RG
dimension, $y_w \approx 0.17$, of the relevant perturbation of the
unstable Ising fixed point.

It is important to observe that the critical behavior for $Q\ge 4$ is
quite peculiar, as the addition of Abelian gauge fields to the spin
(clock) model drives the system out of the unstable spin $XY$ fixed
point towards the related stable $XY_G$ fixed point. To understand
this behavior, it is important to remember that the spin $XY$ and the
$XY_G$ fixed points differ, as duality only relates their thermal
sectors.  In particular, the relevant gauge perturbation of the spin
$XY$ fixed point does not have a counterpart at the $XY_G$ fixed
point, given that the latter is stable---this is our working
hypothesis that is confirmed by the numerical results---when scalar
degrees of freedom are added.

\begin{figure}[tbp]
  \includegraphics*[width=0.9\columnwidth]{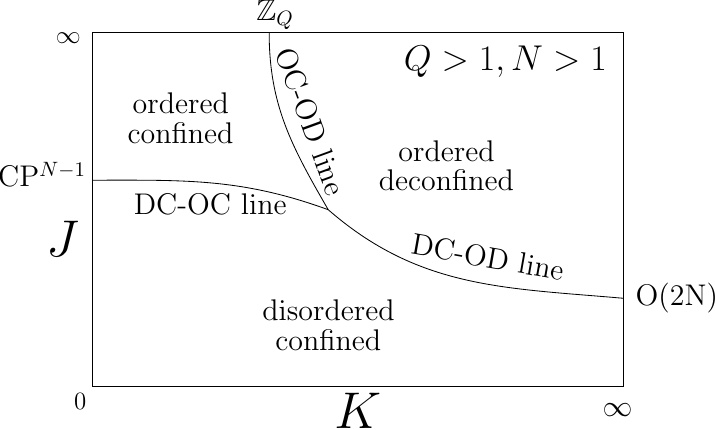}
  \caption{Sketch of the $J$-$K$ phase diagram of the 3D
    multicomponent CLAH model, in which a compact U(1) gauge field is
    coupled with an $N$-component unit-length complex scalar field of
    charge $Q\ge 2$, for generic $N\ge 2$.  There are three phases,
    the disordered-confined (DC), the ordered-deconfined (OD), and the
    ordered-confined (OC) phases.  The AH model is equivalent to the
    $CP^{N-1}$ model for $K=0$, to the $O(2N)$ vector model for
    $K\to\infty$.  For $J\to\infty$ and $Q\ge 2$, we obtain the
    lattice ${\mathbb Z}_Q$ gauge model as in the one-component case.}
\label{phadian}
\end{figure}

As we shall see, the numerical FSS analyses for $Q=2,\,4,\,6$ that we
report in the next section are substantially consistent with the above
scenario. No anomalous violations of universality along the
deconfinement line are observed. Indeed, all deviations can be
explained as standard scaling corrections to the asymptotic critical
behavior.  It is worth mentioning that the noncompact lattice AH
model, which can be seen as the $Q\to\infty$ limit of the CLAH model
~\cite{BPV-22}, has an analogous phase diagram, with an $XY_G$
deconfinement transition line.

We finally mention that a similar analysis has already been performed
for multicomponent CLAH models, in which the scalar field has $N>1$
complex components.  The phase diagram is more complex due to the
possibility of the spontaneous breaking of the global $SU(N)$
symmetry. One observes three different phases, see Fig.~\ref{phadian}:
a low-$J$ disordered confined (DC) phase, and two high-$J$ ordered
phases (OC and OD, respectively) distinguished by the confinement
properties of the gauge field, which is confined for small $K$ and
deconfined for large $K$.  The numerical analyses reported in
Refs.~\cite{BPV-20-hcAH,BPV-22} show that: (i) Along the OC-OD line
transitions are topological and are expected to belong, for any $N$,
to the universality class of the ${\mathbb Z}_Q$ gauge models
(numerical results for $Q=2$ are reported in Ref.~\cite{BPV-20-hcAH});
(ii) The transitions along the DC-OD line are controlled by the
charged fixed point of the AH field theory, and therefore belong to
the same universality class as the transitions between the Coulomb and
Higgs phases of the noncompact lattice AH
model~\cite{BPV-20-hcAH,BPV-21-ncAH,BPV-22,BPV-24}; (iii) Along the
DC-OC line the gauge variables are not critical, so the critical
behavior is expected to be analogous to that of the $CP^{N-1}$ model,
obtained setting $K=0$.

\section{Finite-size scaling behavior of the energy cumulants}
\label{fssenecum}

\subsection{Energy cumulants}
\label{enecu}

Our FSS analyses focus on the gauge-invariant energy cumulants
$C_k$. They can be defined by formally restoring the dependence on the
inverse temperature $\beta$ in Eq.~(\ref{partfun}), as the
inverse-temperature derivatives of the free-energy density, more
precisely
\begin{equation}
C_k=\frac{1}{L^3}\left(\frac{\partial}{\partial\beta}\right)^k
\log Z(\beta,J,K),
\label{Ckdef}
\end{equation}
where $Z$ is the partition function defined in
Eq.~\eqref{partfun}. The cumulants $C_k$ are intensive quantities
which can be related to the energy central moments
\begin{equation}
  M_k = \langle \, (H-\langle H\rangle)^k \, \rangle,
  \label{mkmom}
\end{equation}  
by
\begin{eqnarray}
  & C_1=-L^{-3}\langle H\rangle,\quad   & C_2=L^{-3} M_2, \nonumber \\
  & C_3 = -L^{-3} M_3, \quad   & C_4 = L^{-3} (M_4 - 3 M_2^2),\qquad
\label{Ckdef2}
\end{eqnarray}
etc. Note that $C_2$ is proportional to the specific heat.  These
global quantities allow one to characterize topological transitions in
which no local gauge-invariant order parameter is present, see, e.g.,
Refs.~\cite{SSNHS-03,BPV-20-hcAH,BPV-22-z2g}.

At fixed $K$ and $J$, if the transition is continuous, the cumulant
$C_k$ is expected to show the FSS
behavior~\cite{SSNHS-03,BPV-20-hcAH,BPV-22-z2g}
\begin{eqnarray} 
C_k(\beta J,\beta K,L) &=& {p(J,K)\over q(J,K)^k} L^{k/\nu-d} 
   \Bigl[ {\cal U}_k(Y) + O(L^{-\omega)})\Bigr]  
\nonumber \\ 
&+& B_k(\beta J,\beta K),
  \label{Hg3-scaling-gen}\\
Y &=& {1\over q(J,K)} [\beta - \beta_{c}(J,K)] L^{1/\nu},\label{Ydef}
\end{eqnarray}
where $\beta_{c}(J,K)$ is the critical inverse temperature, $\nu$ is
the length-scale critical exponent, $\omega$ is the leading
correction-to-scaling exponent, and $B_k(\beta J,\beta K)$ is a
regular function---the so-called analytic
background~\cite{PV-02,BPV-20-hcAH}. The functions $p(J,K)$ and
$q(J,K)$ are nonuniversal functions that guarantee that ${\cal
  U}_k(Y)$ is universal, i.e., it is the same in all models that
belong to the same universality class. They are unambiguously defined
only once one fixes their value for a specific $J$ and $K$.  We have
fixed the prefactor in Eq.~(\ref{Hg3-scaling-gen}) so that the
relation $C_{k+1} = \partial C_k/\partial\beta$ implies
\begin{eqnarray}
  {\cal U}_{k+1}(Y) = \partial_Y {\cal U}_k(Y).
  \label{bkrelation}
\end{eqnarray}

We can infer the asymptotic behavior of the scaling
functions ${\cal U}_k(Y)$ for $Y\to\pm \infty$ by matching the FSS
behavior (\ref{Hg3-scaling-gen}) with the leading singular behavior of
$C_k$ in the thermodynamic limit, i.e.,
\begin{equation}
   C_{k,\infty} = \lim_{L\to\infty} C_k \approx c_{k,\pm}
   |\beta-\beta_c|^{-(k-d\nu)} + B_k
\end{equation}
for $\beta\to \beta_c$ at fixed $J$ and $K$. We obtain
\begin{eqnarray}
  {\cal U}_k(Y) \sim |Y|^{-(k-d\nu)}\quad{\rm for}\;\;|Y|\to\infty.
    \label{largeYUk}
\end{eqnarray}
This implies that, if $k - d \nu > 0$, ${\cal U}_k(Y)$ vanishes for
$|Y|\to \infty$, a property than can be used to determine ${\cal
  U}_k(Y)$ from ${\cal U}_{k+1}(Y)$. Indeed, using
Eq.~(\ref{bkrelation}), we can write 
\begin{eqnarray}
  {\cal U}_k(Y) = \int_{-\infty}^Y dZ \,{\cal U}_{k+1}(Z).
  \label{bkintU}
\end{eqnarray}
Taking the limit $Y\to\infty$, we obtain the consistency relation
\begin{equation}
\int_{-\infty}^\infty dY\, {\cal U}_{k+1}(Y) = 0.
\label{intU}
\end{equation}

The cumulants $C_k$ also allow us to define some universal quantities
that are independent of the nonuniversal normalization constants $p$
and $q$ appearing in Eq.~(\ref{Hg3-scaling-gen}).  Since the specific
heat $C_2$ has a single maximum, the corresponding scaling function
${\cal U}_2(Y)$ has a maximum at $Y_{2}$. Moreover, since $C_{3} =
\partial C_2/\partial\beta$, $C_3$ should have a maximum and a minimum
and vanish at the value of $\beta$ where $C_2$ has a maximum.  Thus,
the scaling function ${\cal U}_3(Y)$ should vanish for $Y = Y_{2}$ and
have a maximum and a minimum of opposite sign at two different
values of $Y$, i.e., for $Y=Y_{3,a}$ and $Y = Y_{3,b}$ (in general,
${\cal U}_{k}(Y)$ is expected to have $k-2$ zeroes and $k-1$
stationary points), with $C_{3,\#} = {\cal U}_3(Y_{3,\#})$. We
conventionally identify $Y_{3,a}$ and $Y_{3,b}$ by requiring
$|Y_{3,a}| > |Y_{3,b}|$.  We can then define the universal RG
invariant ratios
\begin{equation}
W_1 \equiv {C_{3,b}\over C_{3,a}}, \qquad
W_2 \equiv {Y_{3,b}\over Y_{3,a}}, \qquad 
W_3 \equiv {Y_{2}\over Y_{3,a}}. \qquad 
\label{Wdef}
\end{equation}

\subsection{Energy cumulants when varying one Hamiltonian parameter}
\label{enecumfixbeta}

Instead of fixing $J$ and $K$ and varying $\beta$, in our numerical
analyses we fix $\beta=1$ and vary $K$ or $J$. Assuming for instance
that $J=\bar{J}$ is fixed, we determine the cumulants $C_k$ defined in
Eqs.~(\ref{Ckdef2}) for different values of $K$. Their FSS behavior is
expected to be analogous to that reported in
Eq.~(\ref{Hg3-scaling-gen}),
\begin{eqnarray} 
&&C_k(\bar{J},K,L) = L^{k/\nu-d} 
   \left[ {\cal C}_k(X) +
  O(L^{-\omega})\right] + b_k(K),\quad
\nonumber \\
&&  X = (K - K_{c}) L^{1/\nu}.\label{Hg3-scaling}
\end{eqnarray}
Comparing with Eq.~(\ref{Hg3-scaling-gen}), we obtain
$b_k(K) = B_k(\bar{J},K)$ and we can relate the scaling functions ${\cal
  C}_k(X)$ with ${\cal U}_k(Y)$.
At fixed $J$ the model has a transition for $K=K_c$, with
$\beta_c(\bar{J},K_c) = 1$. Then, close to criticality we obtain
\begin{equation}
{\cal C}_k(X) = {p(\bar{J},K_c)\over (ac)^k}\ {\cal U}_k (X/a),
\label{Bn-vs-Un}
\end{equation}
with 
\begin{eqnarray}
  Y = {X\over a},\quad a = {q(\bar{J},K_c)\over c},  \quad
  c = \left. 
     -{\partial \beta_c(J,K) \over \partial K} \right|_{\bar{J},K_c}.
\label{Y-Xa}
\end{eqnarray}
As before, we have
\begin{eqnarray}
  {\cal C}_k(X) \approx e_{k,\pm} |X|^{-(k-d\nu)} \quad {\rm for}\;\;X\to\pm \infty,
    \label{largeXBk}
\end{eqnarray}
a result that can be obtained by matching the FSS behavior with the
singular behavior in the thermodynamic limit.  The constants
$e_{k,\pm}$ are related to the constants that parametrize the leading
singular behavior of $C_{k,\infty}(\bar{J},K) = \lim_{L\to\infty}
C_k(\bar{J},K,L)$ for $K\to K_c$:
\begin{equation}
   C_{k,\infty}(K) \approx e_{k,\pm} |K-K_c|^{-(k-d\nu)} + b_k(K).
\end{equation}
If $k - d \nu > 0$, ${\cal C}_k(X)$ vanishes for $|X|\to
\infty$, allowing us to determine ${\cal C}_k(X)$ from 
${\cal
  C}_{k+1}(X)$.  Eqs.~(\ref{bkrelation}) and (\ref{Bn-vs-Un}) imply
${\cal C}_{k+1}(X) = \partial_X {\cal C}_k(X)/ac$, so
write
\begin{eqnarray}
  {\cal C}_k(X) = a c \int_{-\infty}^X dZ \,{\cal C}_{k+1}(Z).
  \label{bkint}
\end{eqnarray}
Taking the limit $X\to\infty$, we obtain the consistency relation
\begin{equation}
\int_{-\infty}^\infty dX\, {\cal C}_{k+1}(X) = 0.
\label{constBK}
\end{equation}

If the universality class of the critical behavior does not change in
the limit $J\to\infty$, therefore for any $Q\neq 4$, we can take the
limit $J\to\infty$. In this case the theory is a function of $\beta K$
only. Thus, if $K_c$ is the critical point of the $J\to\infty$ model
with $\beta = 1$, we have $\beta_c(\infty,K) = K_c/K$, which implies
$c = 1/K_c$. We now use the arbitrariness in the definition of
$p(J,K)$ and $q(J,K)$, setting $p(\infty,K_c) = 1$ and $q(\infty,K_c)
=1/K_c$, thus $a=1$ in Eq.~(\ref{Y-Xa}).  We obtain
\begin{equation}
{\cal U}_k(X) = {1\over K_c^k} {\cal C}_k(X).
\label{UB-Jinf}
\end{equation}
We will use this relation to compute the universal functions ${\cal
  U}_k(X)$ in the Ising case, see App.~\ref{fssz2}. For the $XY_G$
case, we will instead determine ${\cal U}_k(X)$ in the inverted $XY$
gauge model, see App.~\ref{appXY}.

\subsection{Other related quantities}
\label{enecurel}

In the analyses one can also consider cumulants of $H_z$ and
$H_\lambda$, which can be related to derivatives of the free energy
with respect to $J$ or $K$. In particular, we define
\begin{equation}
C_{J,k}=\frac{1}{L^3}\left. \left(\frac{\partial}{\partial J}\right)^k
\log Z \right\vert_{\beta = 1}.
\label{Cjkdef}
\end{equation}
At fixed $J=\bar{J}$ they behave as
\begin{equation}
C_{J,k}(\bar{J},K,L) = L^{k/\nu-d} 
   \left[ {\cal C}_{J,k}(X) +
  O(L^{-\omega})\right] + b_k(K),\quad
\end{equation}
where $X$ is given in Eq.~(\ref{Hg3-scaling}) and $b_k(K)$ is the
analytic contribution. Also ${\cal C}_{J,k}(X)$ can be related to the
universal scaling function ${\cal U}_k(X)$. Indeed, a simple
calculation gives
\begin{equation}
{\cal C}_{J,k}(X) = p(\bar{J},K_c) c_J^k\ {\cal U}_k (X/a),
\label{Jn-vs-Un}
\end{equation}
where $a$ is the same as in Eqs.~(\ref{Y-Xa}) and (\ref{Bn-vs-Un}),
while
\begin{equation}
c_J = - \left. {\partial\over \partial K} 
     \left( {\beta_c(J,K)\over q(J,K)} \right) \right\vert_{\bar{J},K_c} . 
\end{equation}
Of course, analogous equations can be obtained in the case we vary $J$
keeping $K=\bar{K}$ fixed.

\section{Numerical analyses}
\label{numana}

To investigate the nature of the transition lines and check the
predictions outlined in the previous section, we performed MC
simulations, considering cubic lattices of size $L^3$. We use
$C^\star$ boundary conditions~\cite{KW-91, LPRT-16, BPV-21-ncAH},
defined by $\lambda_{{\bm x} + L\hat{\mu},\nu} =
\overline{\lambda}_{{\bm x},\nu}$ and $z_{{\bm x} + L\hat{\mu}} = \bar
z_{\bm x}$, to be able to compare our FSS results with analogous
results obtained in noncompact lattice AH models. We report FSS
analyses for $Q=2,\,4,\,6$.  In particular, for $Q=2$ we discuss in
detail the FSS behavior along the line $K=1$. This is the same region
of phase space considered in Ref.~\cite{SSNHS-03}, which reported
substantial deviations from the Ising asymptotic critical behavior. In
the simulations we perform a combination of overrelaxation
\cite{over1} and Metropolis \cite{metro} updates for the scalar
fields, and Metropolis updates for the gauge fields. For the
Metropolis updates we choose the trial state so as to satisfy detailed
balance and to have an acceptance probability of approximately $30\%$.
For the gauge field we also employ a Metropolis update with proposal
$\lambda_{{\bm x},\mu}\to e^{\pm 2\pi i/Q}\lambda_{{\bm x},\mu}$, in
order to avoid any slowdown of the dynamics in the large-$J$ limit.

In the following we focus on the scaling behavior of the energy
cumulants, testing the RG predictions of Sec.~\ref{fssenecum}.  We
will use the existing estimates of the critical exponents, which are
known with high accuracy both for Ising
~\cite{PV-02,GZ-98,CPRV-02,Hasenbusch-10,KPSV-16,KP-17,FXL-18,Hasenbusch-21}
and $XY$ ~\cite{PV-02, GZ-98, CHPV-06, KP-17, Hasenbusch-19,
  CLLPSSV-20} systems.  In the Ising case, we have $\nu=\nu_I =
0.629971(4)$ and $\omega=\omega_{I} \approx 0.8297(2)$~\cite{KPSV-16}.
For the $XY$ universality class, we have~\cite{CHPV-06,Hasenbusch-19}
$\nu=\nu_{XY}=0.6717(1)$ and $\omega=\omega_{XY}= 0.789(4)$.

The specific heat $C_2$ is not particularly convenient for a numerical
analysis.  Indeed, in the Ising case $C_2$ diverges with the small
exponent $2/\nu_{I}-3= \alpha_I/\nu_I=0.17475(2)$. Therefore, the
leading corrections are due to the background and decay as
$L^{-0.1747}$.  The situation is even worse in the $XY$ case, since
$2/\nu_{XY} - 3 = \alpha_{XY}/\nu_{XY}\approx -0.022$ is negative.  In
this case, the background contribution $b_2$ in FSS
Eq.~(\ref{Hg3-scaling}) is the dominant contribution.

Higher cumulants are more promising. For example, for the third
cumulant we have $3/\nu_I-3 = 1.76235(3)$ and $3/\nu_{XY}-3=
1.4663(7)$ for Ising and $XY$ transitions, respectively.  The leading
scaling corrections decay as $L^{-\omega}$, with $\omega\approx 0.8$
in both cases, so the analytic background is negligible.  However, the
use of higher cumulants also presents some drawbacks. In particular,
statistical errors increase significantly with the order of the
cumulant. Therefore, in the following we only present results up to
the fourth cumulant.

\subsection{Results for $Q=2$}
\label{q2res}

\begin{figure}[tbp]
  \includegraphics*[width=0.9\columnwidth]{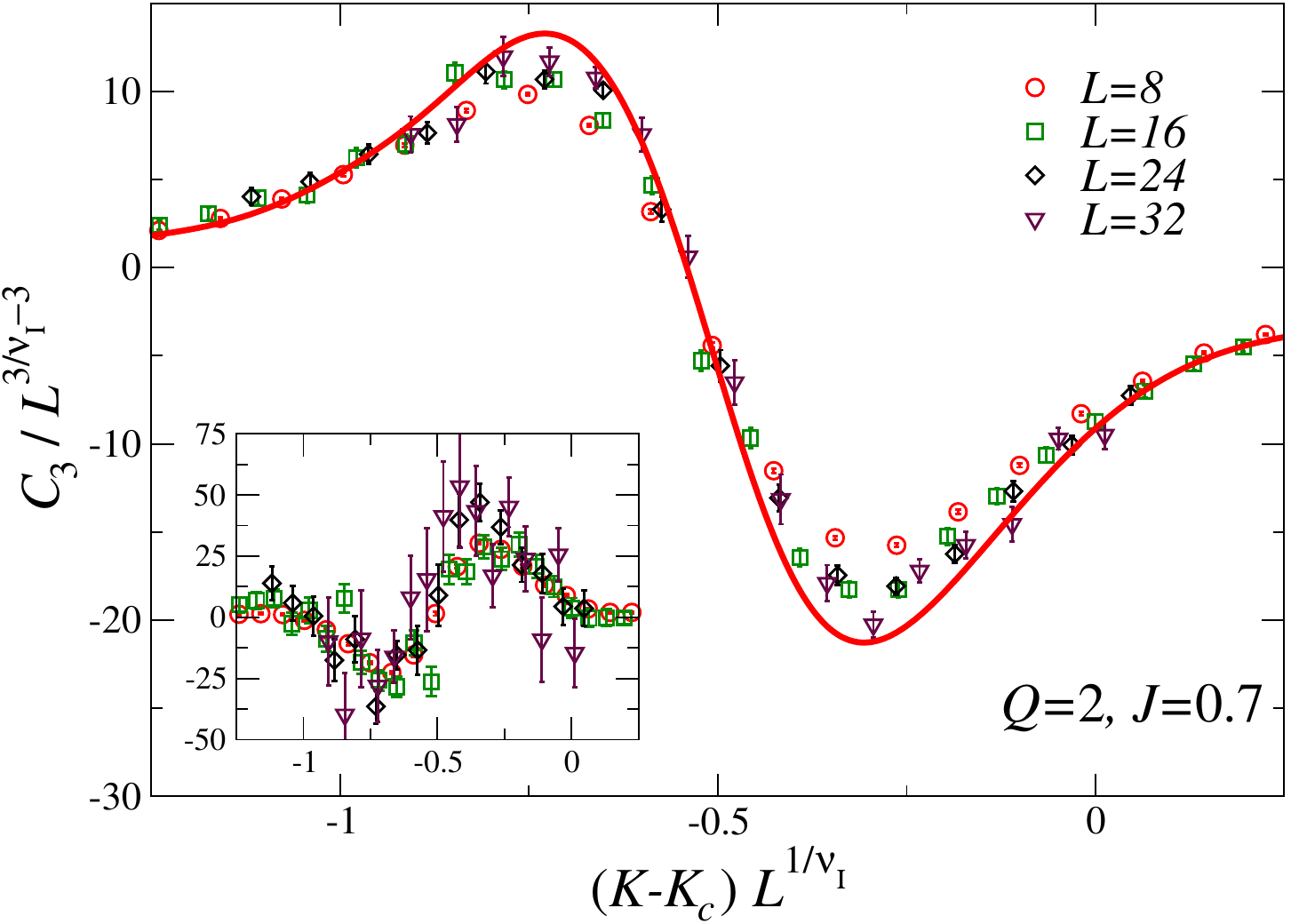}
  \includegraphics*[width=0.9\columnwidth]{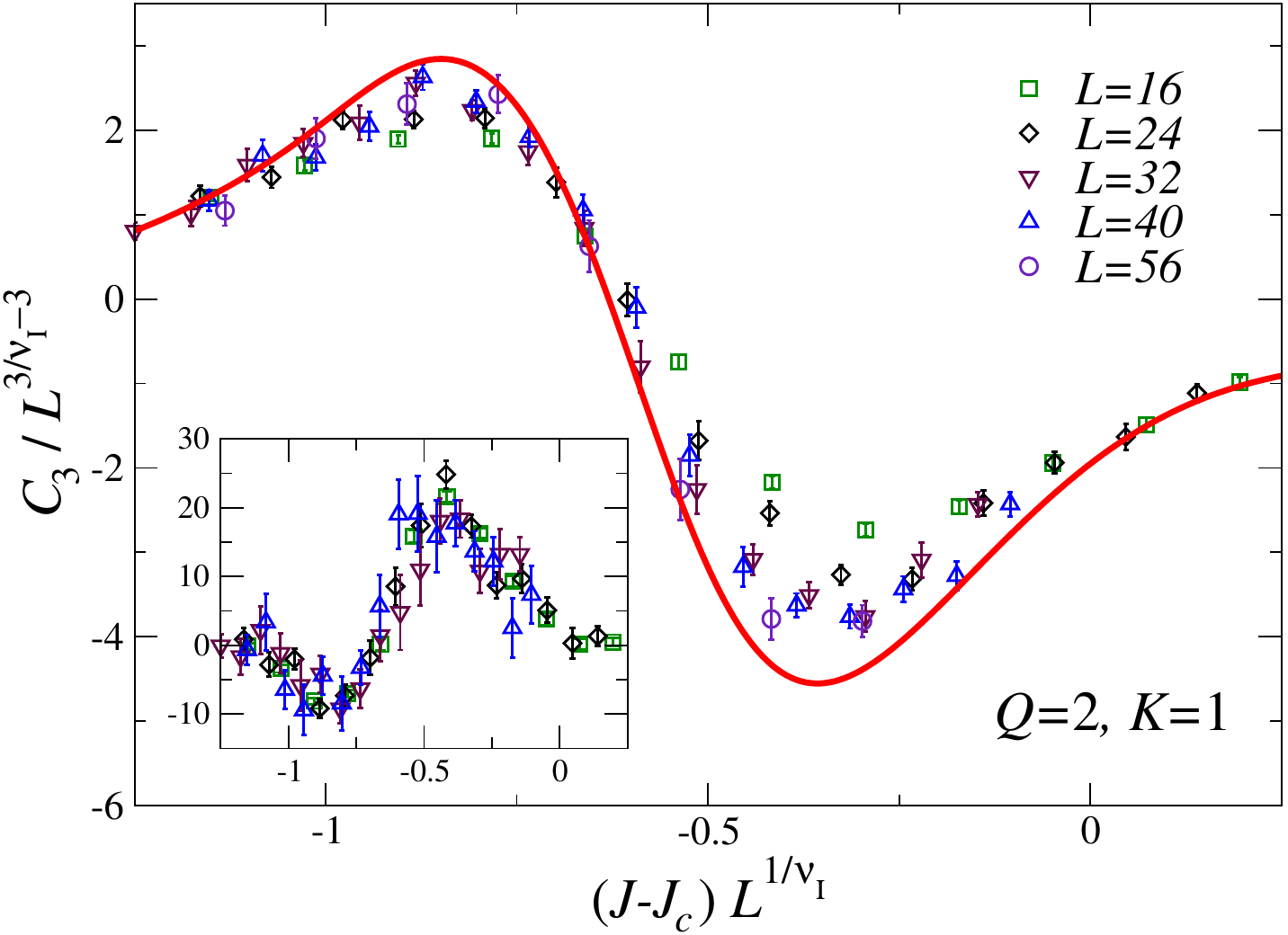}
  \caption{Rescaled $C_3$ cumulant along the $J=0.7$ line (top) and
    the $K=1$ line (bottom) for $Q=2$.  Data approach an asymptotic
    FSS curve, supporting the Ising nature of the transition, although
    significant scaling corrections are visible for $K=1$. The
    continuous curves are computed using Eq.~\eqref{eq:B3B3I}, with $a
    = 0.67$, $p (ac)^{-3} = 2.15$ for $J=0.7$, and with $a = 0.78$, $p
    (ac)^{-3} = 0.46$ for $K=1$; the scaling curve ${\cal
      U}^{(I)}_3(X)$ is reported in App.~\ref{fssz2}.  In the insets
    we plot $L^{\omega_I}\left[ L^{3-3/\nu_I} C_3(J,K,L) - {\cal
        C}_3(X)\right]$ versus $X$ with $\omega_I = 0.83$. The
    observed behavior is consistent with Eq.~(\ref{scacorrB3}), as
    expected for a transition in the Ising universality class.}
\label{q2B3}
\end{figure}

\begin{figure}[tbp]
  \includegraphics*[width=0.9\columnwidth]{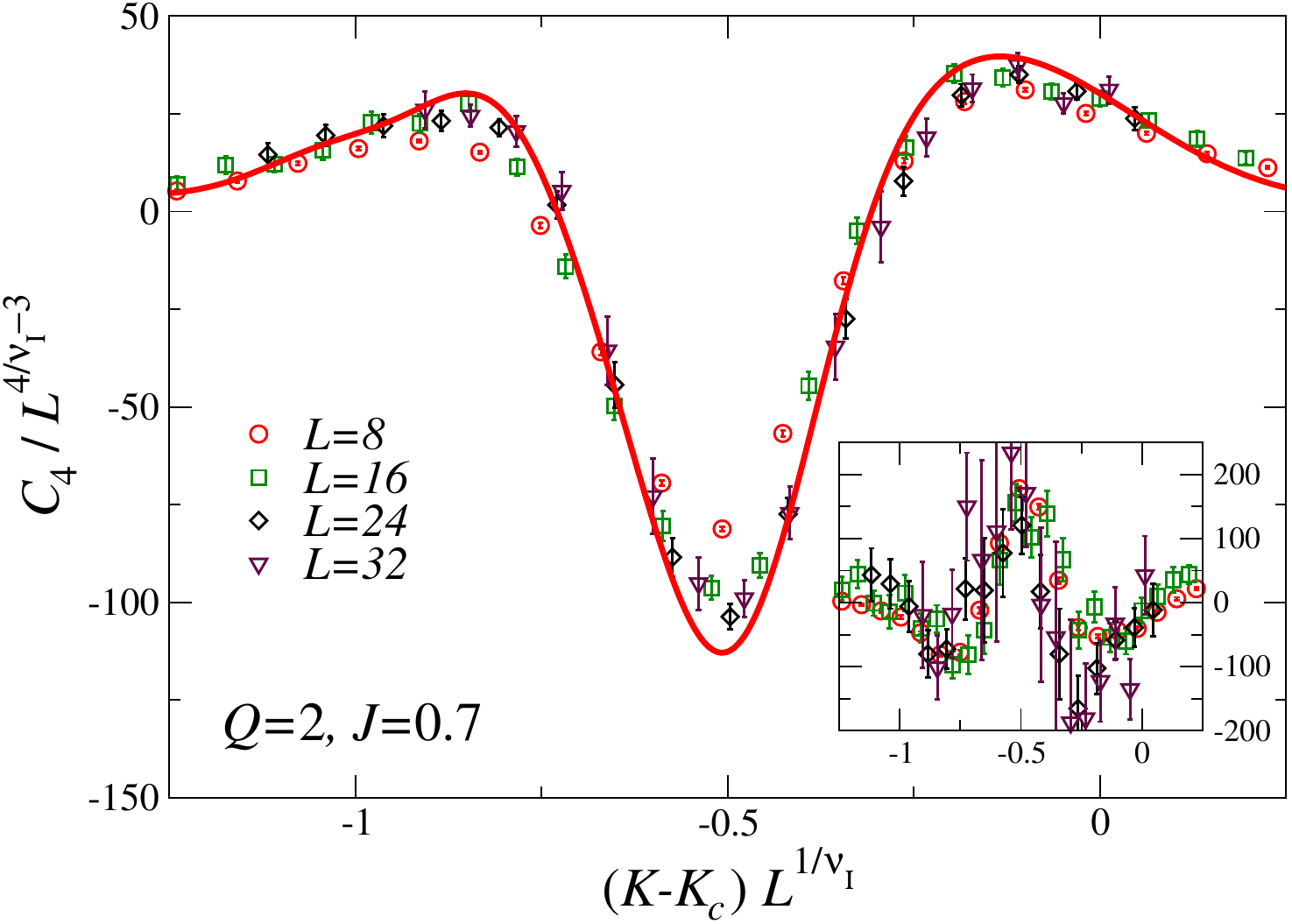}
  \includegraphics*[width=0.9\columnwidth]{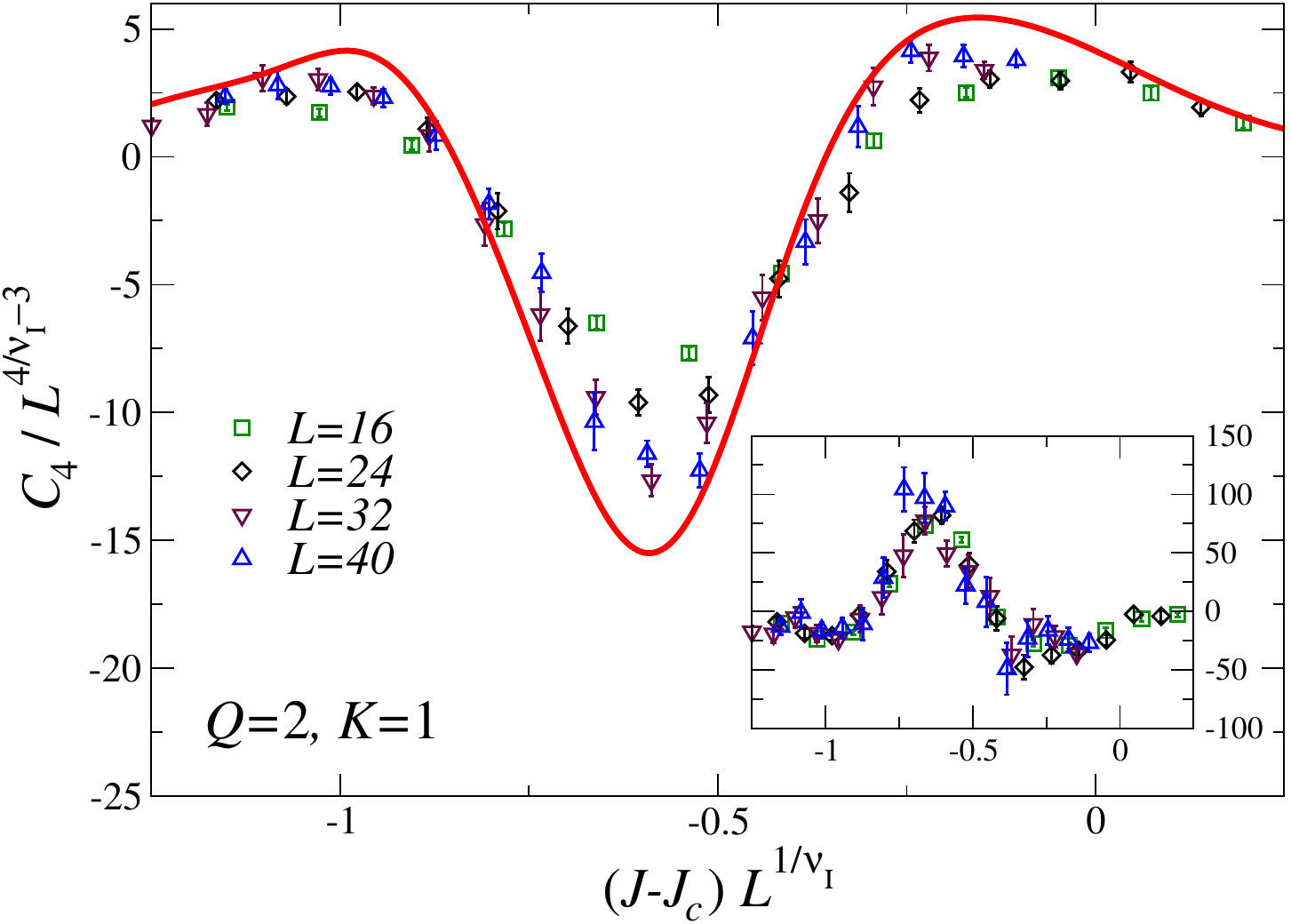}
  \caption{Rescaled $C_4$ cumulant along the $J=0.7$ line (top) and
    the $K=1$ line (bottom) for $Q=2$.  The continuous curves are
    computed using Eq.~\eqref{eq:B3B3I} with $a = 0.67$, $p (ac)^{-4}
    = 2.55$ for $J=0.7$, and with $a = 0.78$, $p (ac)^{-4} = 0.35$ for
    $K=1$. We use the relation ${\cal U}^{(I)}_4(X) = \partial_X {\cal
      U}^{(I)}_3(X)$; the scaling function ${\cal U}^{(I)}_3(X)$ is
    reported in App.~\ref{fssz2}.  We do not report MC data for $K=1$
    and $L=56$ because they are too noisy.  In the insets we plot
    $L^{\omega_I}\left[ L^{3-4/\nu_I} C_4(J,K,L) - {\cal
        C}_4(X)\right]$ versus $X$ with $\omega_I=0.83$. The
    observed behavior is consistent with Eq.~(\ref{scacorrB3}), as
    expected for a transition in the Ising universality class.}
\label{q2B4}
\end{figure}

For $Q=2$ we have performed simulations along the line $K=1$ on
lattices of size up to $L=56$, and along the line $J=0.7$ on lattices
up to $L=32$.  MC estimates of $C_3$ are shown in
Fig.~\ref{q2B3}. They are consistent with the asymptotic Ising-gauge
FSS behavior
\begin{gather}
L^{3-k/\nu_I} C_k(J,K,L) \approx {\mathcal C}_k(X), \label{scalq2}
\\ X=(K-K_c)L^{1/\nu_I} \ \mathrm{or}\ X=(J-J_c)L^{1/\nu_I}.
\end{gather}
However, we note the presence of significant scaling corrections, in
particular, in the data along the $K=1$ line (we will return to this
point later). From fits of $C_3$ to the scaling behavior
(\ref{scalq2}) fixing $\nu_I$ to the Ising value, we obtain
$K_c(J=0.7)=0.5880(2)$ and $J_c(K=1) = 0.34910(6)$. The other
cumulants give compatible, but less precise, results.

A stronger check that the critical behavior is the same as that of the
${\mathbb Z}_2$ gauge model is obtained by comparing the computed
scaling functions ${\cal C}_k(X)$ and the universal curve ${\mathcal
  U}_k^{(I)}$ that we have computed for the $\mathbb{Z}_2$ gauge model
with the same boundary conditions, see
App.~\ref{fssz2}. Eq.~\eqref{Bn-vs-Un} implies
\begin{equation}\label{eq:B3B3I}
\mathcal{C}_k(X)={p\over (ac)^{k}}\,\mathcal{U}_k^{(I)}(X/a),
\end{equation}
where $p$, $a$, and $c$ are constants that depend on the model but not
on the order $k$ of the cumulant. In Fig.~\ref{q2B3}, we compare the
data of the combination $L^{3-3/\nu_I}C_3$ with the scaling curves
$\mathcal{C}_3(X)$ obtained from Eq.~(\ref{eq:B3B3I}). The
normalization constants have been determined to best agreement. We
find $a \approx 0.67$ and $p (ac)^{-3} \approx 2.15$ for $J=0.7$, and
$a \approx 0.78$ and $p (ac)^{-3} \approx 0.46$ for $K=1$, with a
relative uncertainty of approximately 5\%.  In Fig.~\ref{q2B4} we
report the results for $C_4$, together with the scaling curve obtained
by using Eq.~(\ref{Bn-vs-Un}):
\begin{equation}
\mathcal{C}_4(X)={p \over (ac)^{4}}{\mathcal{U}'}_3^{(I)}(X/a),
\end{equation}
where
${\mathcal{U}'}_3^{(I)}(Y)=\mathrm{d}\,{\mathcal{U}}_3^{(I)}(Y)/\mathrm{d}Y$,
see Eq.~\eqref{bkrelation}.  Using the same constant $a$ reported
above, and the estimates $p (ac)^{-4} \approx 2.55$ for $J=0.7$, or $p
(ac)^{-4} \approx 0.35$ for $K=1$ (the error on such estimates is
about 5\% and 10\%, respectively), the Ising-gauge scaling curves are
in reasonable agreement with the numerical data, again confirming the
Ising-gauge nature of the transition. The results obtained from the
analysis of $C_3$ and $C_4$ also allow us to determine the two
constants $p$ and $c$ separately: $p\approx 1.3$, $c\approx 1.3$ for
$J=0.7$ and $p \approx 1.04$, $c\approx 1.7$ for $K=1$.

Another stringent check is obtained by analyzing the deviations of the
finite-volume cumulants from the asymptotic Ising FSS curves. At Ising
transitions, for $k\ge 3$, one generally expects the presence of
corrections vanishing as $L^{-\omega_I}$, with
$\omega_I=0.8297(2)$~\cite{KPSV-16}, see Eq.~\eqref{Hg3-scaling}.
Deviations from the asymptotic curves $\mathcal{C}_k(X)$ are thus
expected to satisfy the relation
\begin{eqnarray}
   L^{\omega_I}\left[ L^{3-k/\nu_I} C_k(J,L) - {\cal C}_k(X)\right]
  \approx {\cal C}_{k,\omega}(X),
  \label{scacorrB3}
\end{eqnarray}
for $k\ge 3$. Note that the scaling functions ${\cal C}_{k,\omega}(X)$
are also expected to be universal apart from an overall rescaling if
one uses $X/a$ as independent variable, where $a$ is the same constant
defined in Eq.~(\ref{eq:B3B3I}). As shown in the insets of
Figs.~\ref{q2B3} and \ref{q2B4}, both $C_3$ and $C_4$ data are nicely
consistent with Eq.~(\ref{scacorrB3}) and with the universality (apart
from a rescaling) of ${\cal C}_{k,\omega}(X)$ when expressed in terms
of $X/a$ (note that $a$ is approximately the same for the data along
the two different lines). Eq.~(\ref{scacorrB3}) does not hold for
$C_2$, since the leading corrections are due to the analytic
background and decay as $L^{-\alpha_I/\nu_I}$, with
$\alpha_I/\nu_I\approx 0.1747$.

\begin{figure}[tbp]
  \includegraphics*[width=0.9\columnwidth]{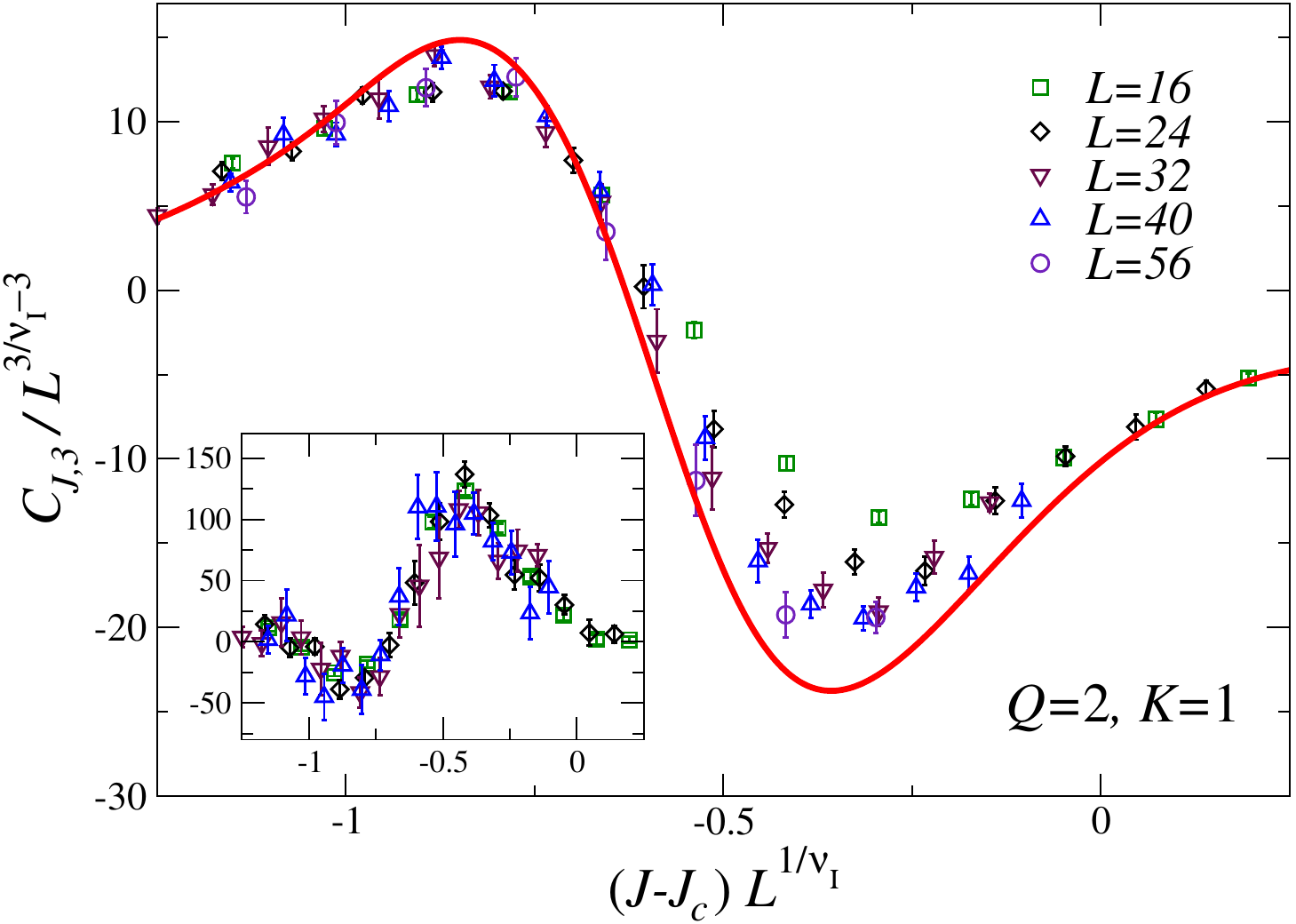}
  \includegraphics*[width=0.9\columnwidth]{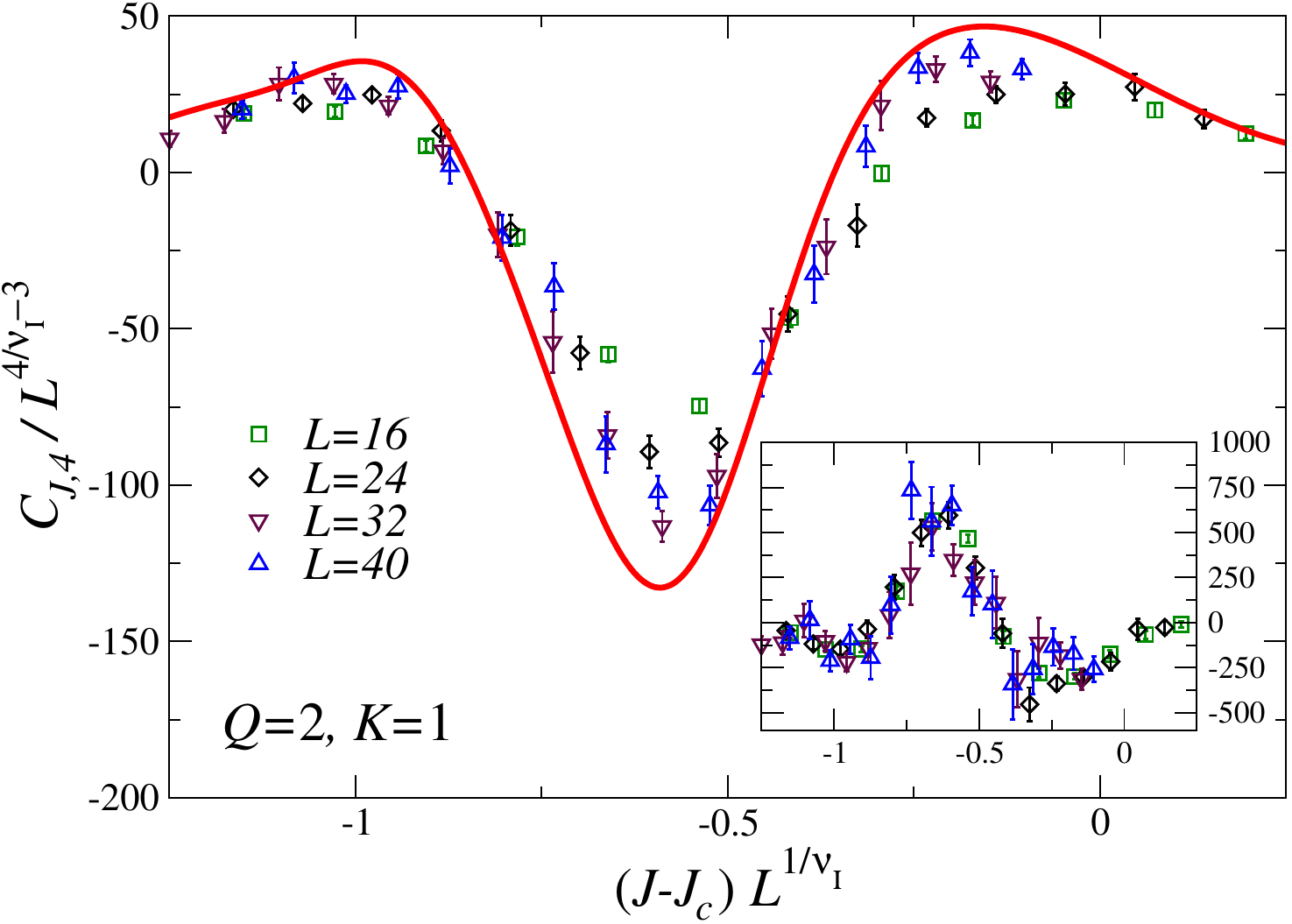}
  \caption{Results for $Q=2$ along the line $K=1$.  Rescaled cumulants
    $C_{J,3}$ (top) and $C_{J,4}$ (bottom), for $Q=2$ along the line
    $K=1$, using the Ising critical exponent $\nu_I = 0.629971$.  The
    continuous curves are computed using Eq.~\eqref{Jn-vs-Un}, with $a
    = 0.78$, $p = 1.04$, $c_J=0.97$; the scaling curve ${\cal
      U}^{(I)}_3(X)$ is reported in App.~\ref{fssz2}.  We do not
    report estimates of $C_{J,4}$ for $K=1$ and $L=56$ because data are
    too noisy.  In the insets we plot the combination defined as in
    Eq.~(\ref{scacorrB3}) versus $X$ with $\omega_I= 0.83$.  }
\label{q2K1bis}
\end{figure}

We also note that the value $J_c\approx 0.349$ obtained for $K=1$ is
close to the asymptotic $K\to\infty$ critical value $J_{XY}\approx
0.227$, indicating that the transition point at $K=1$ is already in a
region where the transition line runs almost parallel to the $K$-axis
in Fig.~\ref{phadia}. In this situation it might be more natural to
use the cumulants $C_{J,k}$ of $H_z$ instead of the cumulants $C_k$
defined in Eq.~\eqref{Ckdef}. Estimates of $C_{J,3}$ and $C_{J,4}$ for
$K=1$ are reported in Fig.~\ref{q2K1bis}, together with the scaling
curves obtained using Eq.~(\ref{Jn-vs-Un}). Since the constant $p$ has
already been determined using the cumulants $C_k$, only the constant
$c_J$ can be tuned.  If we set $c_J = 0.97$, we observe a reasonable
agreement for both $C_{J,3}$ and $C_{J,4}$. Finally, we analyzed the
scaling corrections, considering the combination defined as in
Eq.~(\ref{scacorrB3}). Data scale nicely, see the insets in
Fig.~\ref{q2K1bis}. Note also that, as predicted by the RG theory, the
scaling curves of the finite-volume corrections differ only by a
multiplicative nonuniversal factor from those computed using the
cumulants $C_3$ and $C_4$, see the insets of Figs.~\ref{q2B3} and
\ref{q2B4} for $K=1$.

The above results show that the critical behavior of the energy
cumulants for $Q=2$ and $K=1$ is fully compatible with the Ising-gauge
nature of the transition.  They are in apparent contradiction with the
numerical analysis of Ref.~\cite{SSNHS-03}, which reported significant
deviations from the Ising behavior in the same region of the phase
diagram (see Fig. 7 of Ref.~\cite{SSNHS-03}, keeping into account that
the ratio $\kappa/\beta$ in Ref.~\cite{SSNHS-03} corresponds to $K/J$
in our notations, thus $\kappa/\beta\approx 2.9$ at the transition
point for $K=1$).

We also remark that, with increasing $K$, the regime where the
eventual asymptotic FSS behavior can be observed is expected to be
pushed toward larger and larger values of $L$, due to the emergence of
a preasymptotic regime in which the critical behavior is controlled by
the $XY$ spin fixed point.  Thus, when the correlation length $\xi$ of
the critical modes, or the lattice size $L$ in finite-size systems, is
smaller than the crossover length $\xi_\alpha\sim K$, we expect to
observe an apparent $XY$ critical behavior because the gauge modes are
effectively frozen. These preasymptotic crossover effects can be
hardly disentangled when performing FSS analyses with data in a
limited range of lattice sizes, and may lead to apparent variations of
the effective critical exponents with $L$.  These crossover effects do
not allow us to analyze the critical behavior for values of $K$
significantly larger than 1. For instance, for $K=2$ one would
probably need precise results on lattices of size up to $L=100$ to
observe the asymptotic behavior with reasonable precision.  These
simulations would require a huge numerical effort.\footnote{In our
simulations for $Q=2$ and $K=1$ a statistics of the order of $10^{8}$
lattice updates was gathered for each value of $L$, requiring about
$10^5$ core-hours of total CPU time.}

In conclusion, the above FSS analyses show that the deconfinement
transitions in the $Q=2$ CLAH model are consistent with the expected
Ising critical behavior. They belong to the same universality class as
the transition in the lattice ${\mathbb Z}_2$ gauge model. All
violations of universality are explained by standard scaling
corrections, as predicted by the RG theory of critical
phenomena~\cite{PV-02}.  We expect substantial crossover phenomena for
large values of $K$, when approaching the $XY$ transition at
$K=\infty$, which may be naively interpreted as violations of
universality in numerical analyses.  Therefore, our results do not
confirm the existence of a fixed-point line with varying critical
exponents put forward in Ref.~\cite{SSNHS-03}.

\subsection{Results for $Q=6$}
\label{q6res}

\begin{figure}[tbp]
  \includegraphics*[width=0.9\columnwidth]{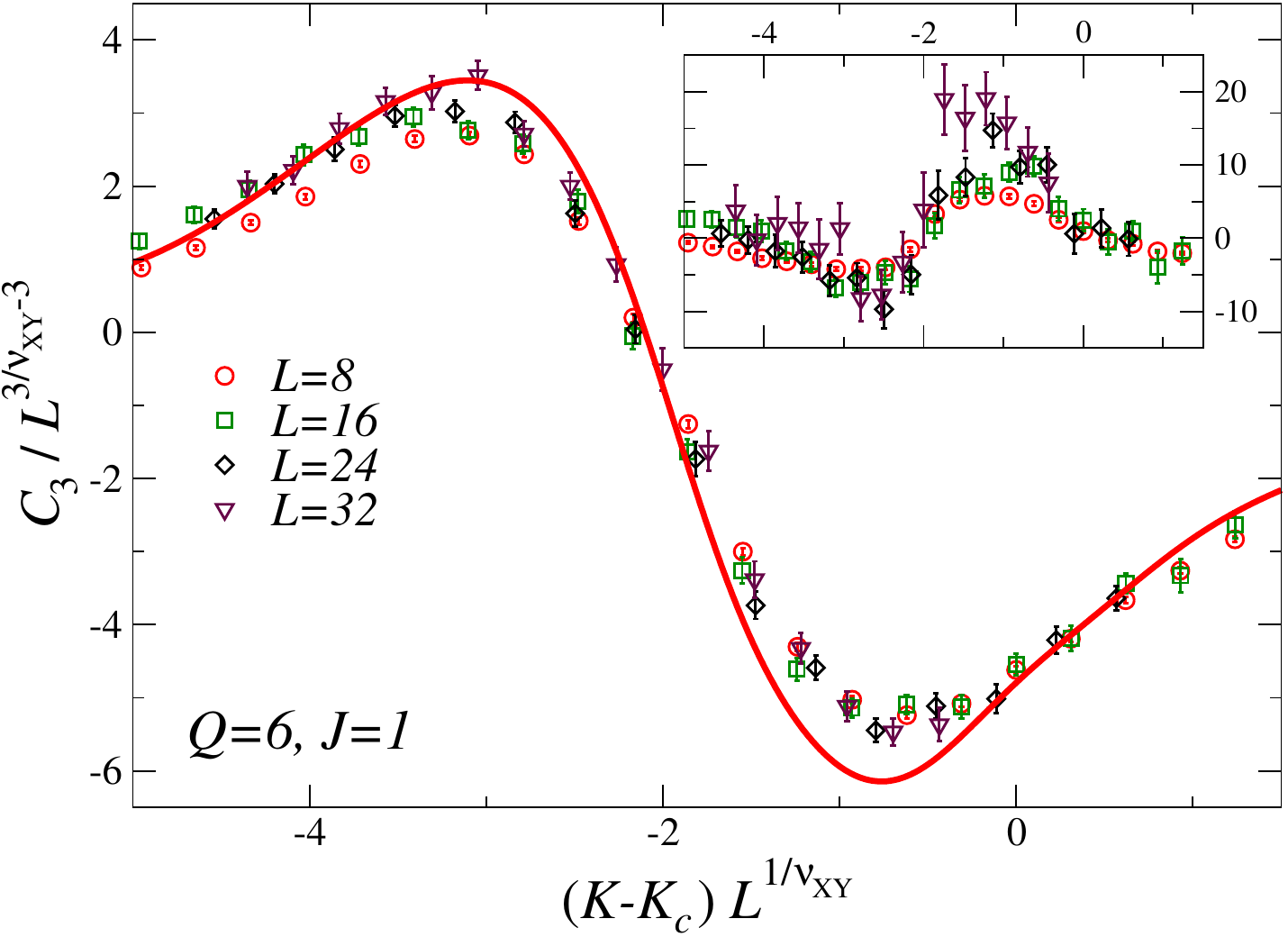}
  \caption{FSS behavior of the third cumulant $C_3$ for $Q=6$ along
    the line $J=1.0$, using the $XY$ critical exponent $\nu_{XY} =
    0.6717$.  We also report (solid curve) the scaling curve
    (\ref{eq:B3B3IXY}), obtained by using the parametrization of
    $\mathcal{U}_3^{(XY)}(X)$ reported in App.~\ref{appXY}.  In the
    inset we plot $L^{\omega_{XY}}\left[ L^{3-3/\nu_{XY}} C_3(J,L) -
      {\cal C}_3(X)\right]$, with $\nu_{XY} = 0.6717$ and $\omega_{XY}
    = 0.789$. }
\label{q6}
\end{figure}

We have performed an analogous FSS analysis of the energy cumulants
for $Q=6$ at fixed $J=1.0$. In this case, see the general arguments
reported in Sec.~\ref{phasediagram}, we expect the deconfinement
transitions to belong to the $XY_G$ universality class.  This is
related to the fact that, for $Q>4$, the $J\to\infty$ limit of the
model, the lattice ${\mathbb Z}_{Q}$ gauge model, undergoes a
topological phase transition in the $XY_G$ universality class.  $XY_G$
transitions are expected along the whole transition line, as it occurs
in the one-component noncompact lattice AH model~\cite{BPV-24}.

As already remarked in Sec.~\ref{fssenecum}, the behavior of the
second cumulant $C_2$, which corresponds to the specific heat, is
dominated by the analytic background. Therefore, we focus on the third
cumulant. Results along the line $J=1.0$ are shown in
Fig.~\ref{q6}. They clearly support the expected $XY_G$ behavior. The
estimates of $L^{3-3/\nu_{XY}} C_3$ plotted versus
$X=(K-K_c)L^{1/\nu_{XY}}$ appear to approach a scaling function ${\cal
  C}_3(X)$ if we use the $XY$ value $\nu_{XY} =0.6717$ and $K_c \approx
2.0808(4)$. In Fig.~\ref{q6} we also report (solid curve)
\begin{equation}\label{eq:B3B3IXY}
  \mathcal{C}_3(X)={p\over (ac)^{3}} \,\mathcal{U}_3^{(XY)}(X/a),
\end{equation}
where $a\approx 35$, $p (ac)^{-3}\approx 7.5\cdot 10^{-5}$ 
(with a relative uncertainty of about 10\%), and
$\mathcal{U}_3^{(XY)}(X)$ is the FSS scaling function of the $XY_G$
universality class computed with the same boundary conditions, see
App.~\ref{appXY}.  Deviations from the asymptotic curve are expected
to scale as in Eq.~(\ref{scacorrB3}) (with $\nu_{XY}$ replacing
$\nu_I$).  The results, shown in the inset of Fig.~\ref{q6}, are
consistent with the $XY_G$ nature of the transition.  We do not
present data for the cumulant $C_4$, because the available data are
too noisy.

\begin{figure}[tbp]
  \includegraphics*[width=0.9\columnwidth]{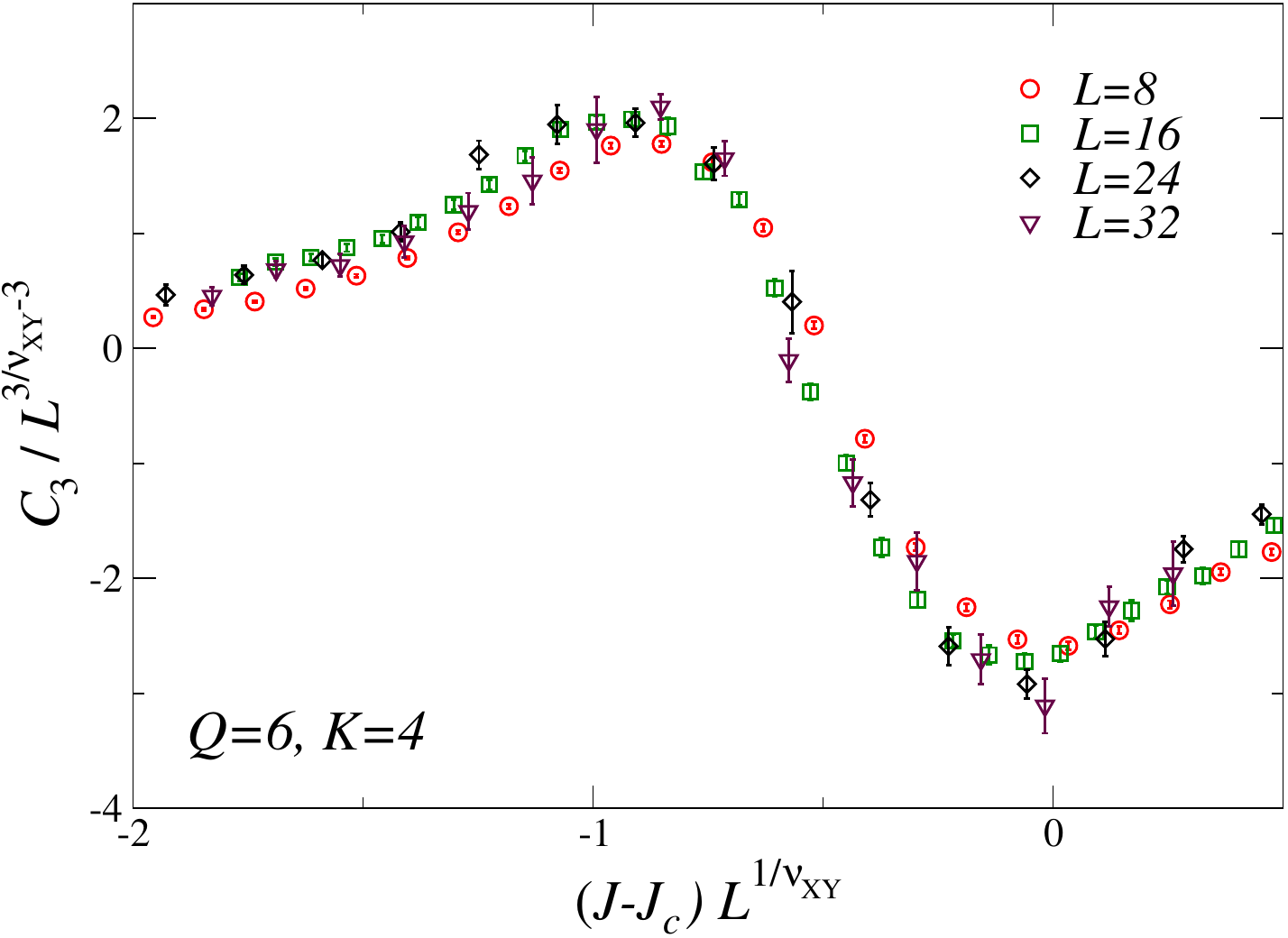}
  \caption{FSS behavior of the third cumulant $C_3$ in the $Q=6$ model
    along the $K=4$ line, with $\nu_{XY}= 0.6717$ and $J_c=0.4684$.
  }
\label{Q6K4}
\end{figure}

We also performed some MC simulations along the line $K=4$. We observe
a transition at $J_c= 0.4684(3)$ with exponents consistent with those
of the $XY_G$ universality class. The scaling curve of $C_3$ shows
some apparent deviations from the one obtained using
Eq.~(\ref{eq:B3B3IXY}), for any choice of the nonuniversal parameters
$a$, $p$, and $c$, see Fig.~\ref{Q6K4}.  This can be again explained
as a crossover effect due to the presence of the unstable spin $XY$
fixed point. Indeed, given that the $XY$ and $XY_G$ universality
classes differ, we expect the scaling curves computed along the
deconfinement line to be unrelated with those computed in the $XY$
spin model with the same boundary conditions (see App.~\ref{App.B} for
a discussion of the appropriate boundary conditions). For large $K$
and finite $L$ we expect data to show a crossover, with an apparent
behavior related to that observed in the spin model for small values
of $L$ and, more precisely, for $L\ll \xi_\alpha$, where $\xi_\alpha$
is the crossover length.

\begin{figure}[tbp]
  \includegraphics*[width=0.9\columnwidth]{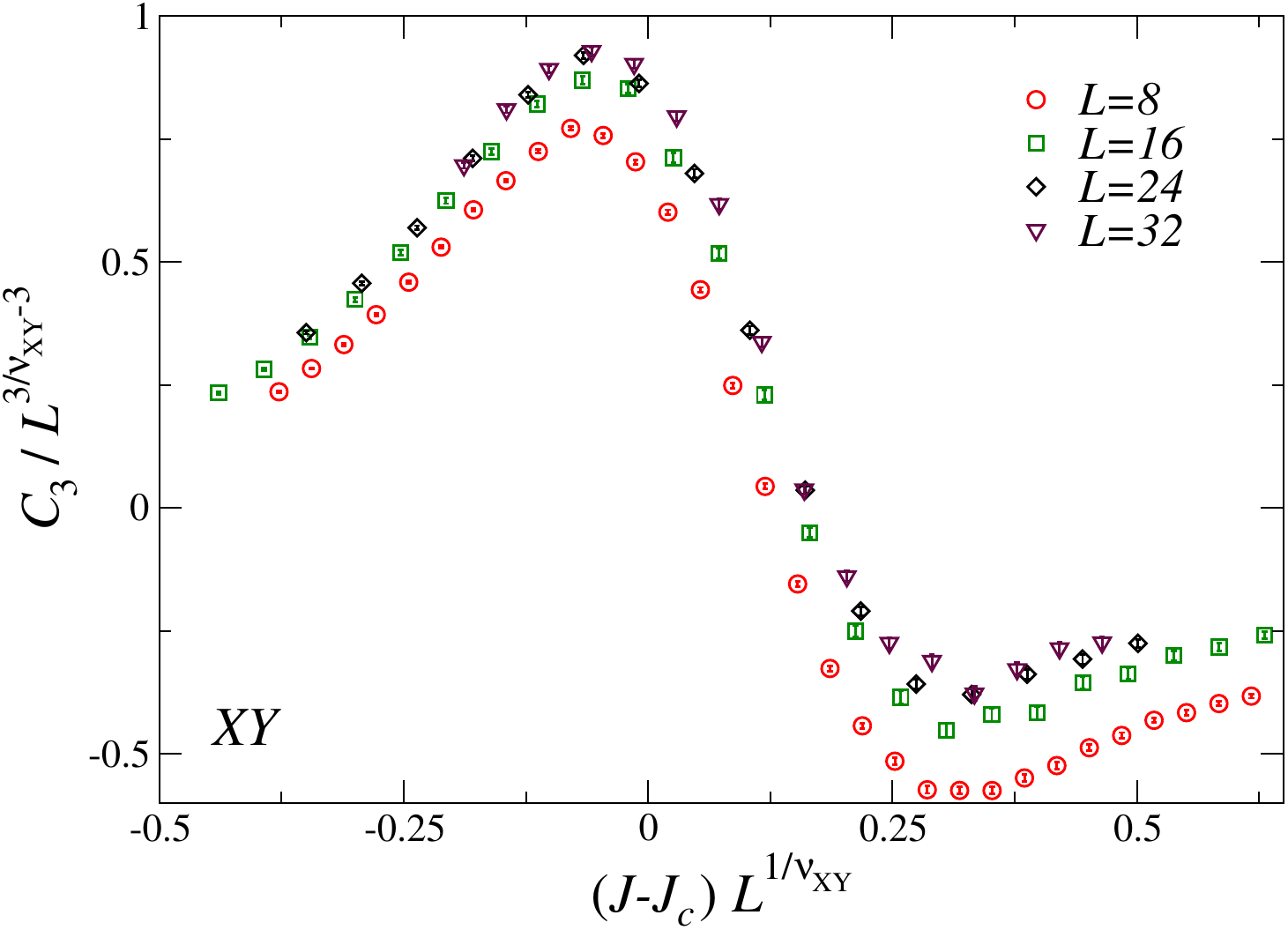}
  \caption{FSS behavior of the third cumulant $C_3$ in the $XY$ vector
    model, with $\nu_{XY}= 0.6717$ and $J_c=0.22708234$.  }
\label{XY}
\end{figure}

To clarify this point we have determined the cumulants in the $XY$
spin model with analogous boundary conditions.  Results are shown in
Fig.~\ref{XY}.  Although significant scaling corrections are present,
nevertheless, some qualitative features clearly distinguish the spin
$XY$ scaling curve in Fig.~\ref{XY} from that of the $XY_G$
universality class, see Fig.~\ref{q6}.  For instance, in the spin $XY$
case the maximum and the minimum are located on opposite sides of the
critical point. Instead, they are located on the same side in the
$XY_G$ case.

The results presented in Fig.~\ref{Q6K4} show that the FSS behavior of
$C_3$ at $K=4$ is somewhat intermediate between that expected for the
vector $XY$ and for the $XY_G$ universality classes.  Again, this can
be understood by looking at the position of the minimum, which, for
$q=6$ and $K=4$ almost coincides with the critical-point position, see
Fig.~\ref{Q6K4}. For a quantitative comparison we report the ratio
$W_2$ defined in Eq.~(\ref{Wdef}): we have $W_2 \approx 0$ for the
$Q=6$ model along the line $K=4$, $W_2 \approx -0.2$ for the $XY$
model, and $W_2 = 0.25(5)$ for the $XY_G$ universality class. We can
also compare the ratio $W_1$ that is related to the height of the
peaks. For this quantity, the $Q=6$ data at $K=4$ are quite different
from the $XY$ ones and close to the $XY_G$ results.  We find $W_1
\approx -1.5$ for the $Q=6$ model at $K=4$ on the lattices studied, to
be compared with $W_1\approx -2.4$ and $W_1 = -1.75(4)$ for the $XY$
($L=32$ lattice) model and the $XY_G$ universality class,
respectively.

\subsection{Results for $Q=4$}
\label{q4res}

\begin{figure}[tbp]
  \includegraphics*[width=0.9\columnwidth]{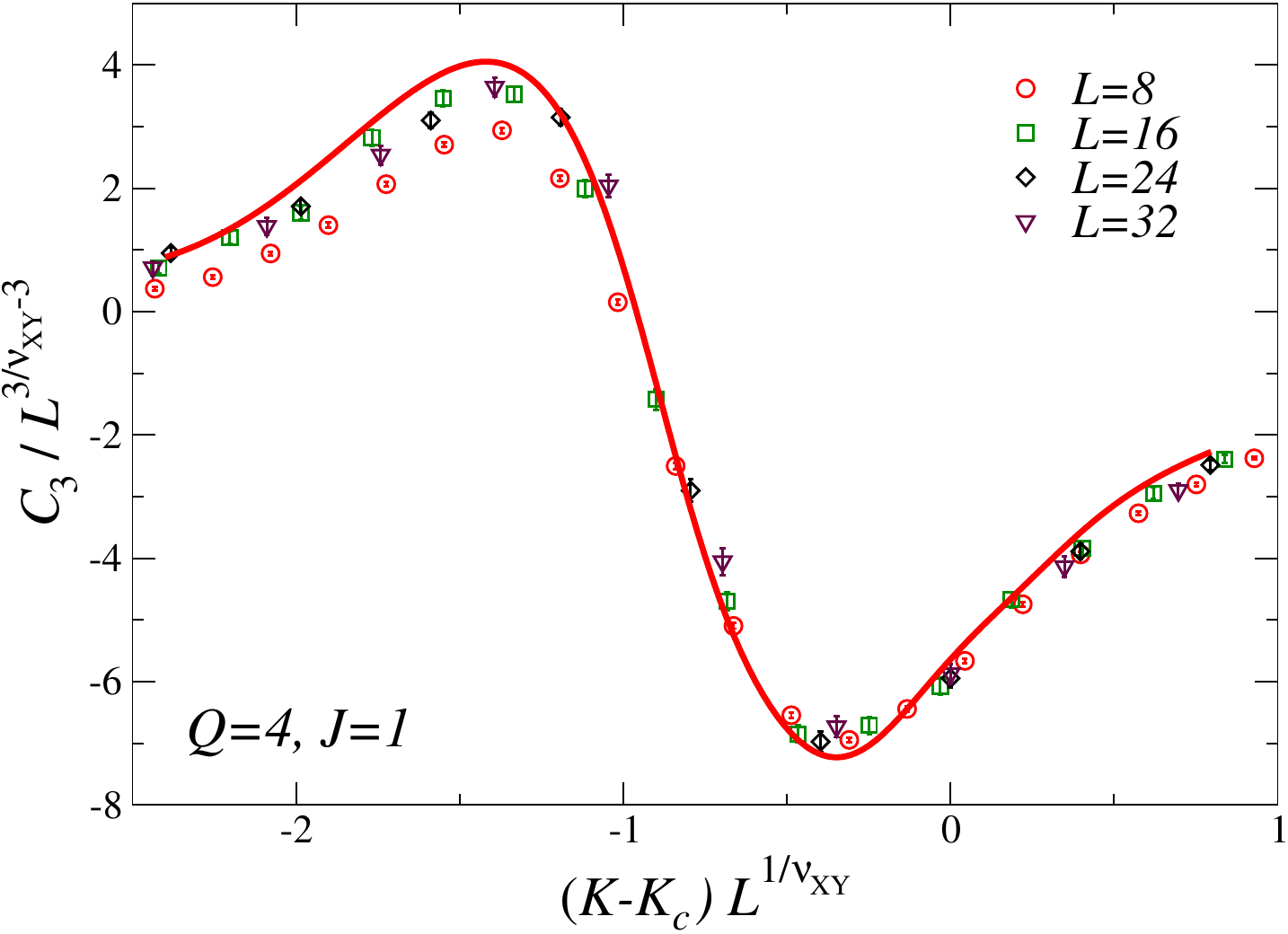}
  \caption{FSS behavior of the third cumulant $C_3$ for $Q=4$ along
    the line $J=1.0$, with $\nu_{XY} = 0.6717$, $J_c = 1.0205$. Data
    approach the FSS scaling curve computed using
    Eq.~(\ref{eq:B3B3IXY}); we use the parametrization of
    $\mathcal{U}_3^{(XY)}(X)$ reported in App.~\ref{appXY}, $a = 16$ ,
    $p (ac)^{-3} = 9\cdot 10^{-5}$.  }
\label{q4}
\end{figure}

Finally, we present some results for $Q=4$.  As discussed in
Sec.~\ref{phasediagram}, this model is expected to represent an
exception with respect to the general scenario. Indeed, along the
deconfinement line transitions are expected to belong to the $XY_G$
and not to the Ising universality class, that characterizes the
critical behavior of the lattice ${\mathbb Z}_4$ model obtained in the
$J\to\infty$ limit.  This prediction is confirmed by the analysis of
the numerical data obtained along the line $J=1$.  In Fig.~\ref{q4} we
report a scaling plot of $C_3$ using the $XY$ value $\nu_{XY} =
0.6717$ and $K_c\approx 1.0205(2)$. We also show the scaling curve
obtained by using Eq.~\eqref{eq:B3B3IXY}, using the $XY_G$ scaling
curve with $a=16$ and $p (ac)^{-3} = 9\cdot 10^{-5}$ (with a relative
uncertainty of about 5\%).  Data clearly confirm the $XY_G$ nature of
the transition.  We also mention that the MC data are in clear
disagreement with the alternative Ising scenario.

\section{Conclusions}
\label{conclu}

We have investigated the nature of the deconfinement transitions of
the one-component multicharge CLAH models defined in
Eqs.~(\ref{partfun})-(\ref{plaquette}) (see Fig.~\ref{phadia} for a
sketch of the phase diagram). We argue that they belong to the same
universality class as the continuous transitions in generic ${\mathbb
  Z}_Q$ lattice gauge models without scalar fields. In particular, the
transitions belong to the Ising-gauge and $XY_G$ universality classes
for $Q=2$ and $Q\ge 4$, respectively.  For $Q=3$ we expect first-order
deconfinement transitions, since there is no universality class with
${\mathbb Z}_3$ gauge symmetry.  This scenario contrasts with the one
put forward in Ref.~\cite{SSNHS-03}, which suggested the existence of
a fixed-point line with continuously varying critical exponents.

To accurately investigate the transitions of the single-component
model, we consider the FSS behavior of the energy cumulants, see,
e.g., Eq.~(\ref{Ckdef2}). For these quantities we derive several
useful general properties that can be used to identify the nature of
the critical behavior in any topological transition. We apply these
general techniques to the CLAH model with $Q=2,4,6$ for several values
of the Hamiltonian parameters.
 
Our FSS analyses for $Q=2$ indicate that the continuous transitions
along the deconfinement transition line belong to the Ising-gauge
universality class, i.e., that the critical behavior is the same as in
the lattice ${\mathbb Z}_2$ gauge model, which is formally obtained in
the $J\to\infty$ limit. We remark that this check requires a careful
analysis of the nonuniversal scaling corrections, which are predicted
by the RG approach.  Indeed, in FSS analyses of data obtained for
relatively small lattices, irrelevant scaling fields or nearby
unstable fixed points may give rise to apparent violations of
universality. For instance, significant crossover effects (with a
crossover length $\xi_\alpha\sim K$) are expected for large values of
$K$, due to the $XY$ transition occurring for $K=\infty$.

For $Q\ge 4$, the deconfinement transitions should belong to the
$XY_G$ universality class, with an effective enlargement of the
symmetry from ${\mathbb Z}_Q$ to $U(1)$.  This is supported by the FSS
analyses of the energy cumulants for $Q=4$ and $Q=6$. Note that for
$Q=4$, at variance with what occurs for $Q>4$, the critical behavior
along the deconfinement transition line differs from that observed in
the ${\mathbb Z}_4$ gauge model obtain for $J\to\infty$, which instead
belongs to the Ising-gauge universality class.  Crossover phenomena
are expected for large values of $K$ also for $Q\ge 4$.  This
crossover is quite peculiar as the unstable fixed point is the spin
$XY$ fixed point, while the stable one is the gauge $XY_G$ fixed
point. The two fixed points are different, although related by
duality. This implies that, even though critical exponents are the
same, other universal properties (for instance, the FSS functions)
differ.

We finally mention that, for $Q=3$, the CLAH model is expected to
undergo first-order transitions for any finite $K$ and $J$.  However,
since the continuous $XY$ transition is approached for $K\to\infty$,
we expect that the first-order transitions become weaker as $K$
increases. More precisely, the first-order nature becomes apparent
only for $L\gg \xi_\alpha$, where $\xi_\alpha\sim K$ is the crossover
length.  For smaller sizes, sizable crossover effects are expected,
with an apparent $XY$ critical behavior. These crossover effects may
explain the results of Ref.~\cite{SSNHS-03}, which apparently observed
a change of the nature of the transition with increasing $K$, from a
first-order to a continuous one.

\begin{figure}[tbp]
\includegraphics*[width=0.95\columnwidth]{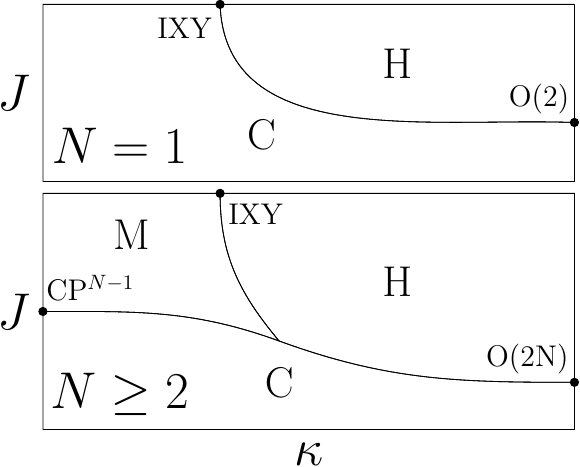}
\caption{The phase diagram of the $N$-component nCLAH model
  (\ref{nCLAHham}), in the Hamiltonian parameter space $\kappa$-$J$,
  for $N=1$ (top) and generic $N\ge 2$ (bottom).  For $N=1$, there are
  two phases, the Coulomb (C) and Higgs (H) phases, characterized by
  the confinement and deconfinement of charged gauge-invariant
  excitations, respectively.  For $N\ge 2$, the scalar field is
  disordered and gauge correlations are long ranged in the small-$J$
  Coulomb (C) phase.  For large $J$ two phases occur, the molecular
  (M) and Higgs (H) ordered phase, in which the global $SU(N)$
  symmetry is spontaneously broken.  The two phases are distinguished
  by the behavior of the gauge modes: the gauge field is long ranged
  in the M phase (small $\kappa$), while it is gapped in the H phase
  (large $\kappa$).  Moreover, while the C and M phases are confined
  phases, the H phase shows the deconfinement of charged
  gauge-invariant excitations.  See, e.g., Ref.~\cite{BPV-24} for more
  details.}
\label{phadianc}
\end{figure}

It is worth looking at these results in a more general context,
considering multicomponent compact and noncompact lattice AH models.
To begin with, let us compare the phase diagrams of the one-component
($N=1$) with those of the multicomponent ($N\ge 2$) CLAH models,
reported in Figs.~\ref{phadia} and \ref{phadian}, respectively.  In
the multicomponent model there are three phases, separated by three
transition lines. The topological OC-OD transition line, which
separates two ordered phases, is the analogue of the deconfinement
line in the one-component model. Also in multicomponent models the
scalar fields do not directly contribute to the critical behavior
along such transition line, which belongs to the ${\mathbb Z}_Q$ gauge
universality class~\cite{BPV-20-hcAH}.  The other two lines (DC-OC and
DC-OD) are instead associated with the spontaneous breaking of the
$SU(N)$ symmetry and are therefore specific of the multicomponent CLAH
models.  Along, the DC-OC line only scalar fields are critical, and
thus the transition is the same as in the $CP^{N-1}$ model.  Instead,
the DC-OD transitions are driven by a nontrivial interplay between the
$U(1)$ gauge and $SU(N)$ global symmetry. For $N < N^*$, $N^*=7(2)$
these transitions are of first order. For $N\ge N^*$ the DC-OD
transitions can be continuous, controlled by the stable fixed point of
the multicomponent AH field theory~\cite{BPV-22,BPV-20-hcAH}.

It is also interesting to compare the phase diagrams of the CLAH
models with those of the noncompact lattice AH (nCLAH) models (see
App.~\ref{appXY} for definitions) reported in Fig.~\ref{phadianc},
which present various phases and transition lines (see, e.g.,
Refs.~\cite{BPV-24,BPV-21-ncAH} for more details).  The topology of
the phase diagrams is similar both in the one-component and in the
multicomponent case. This is not surprising since, as argued in
Ref.~\cite{BPV-22}, the nCLAH models can be obtained as an appropriate
large-$Q$ limit of the CLAH models.  Interestingly, also the critical
behavior in compact and noncompact models is the same, except in one
case we discuss below.  In multicomponent models the behavior along
the CM transition line of the nCLAH model is the same as along the
DC-OC line of the CLAH models for any $Q\ge 2$. The same is true for
the $Q=1$ multicomponent CLAH models~\cite{PV-19-AH3d}, in which there
is a single transition line, which has the same nature as the CM
transitions of the nCLAH models and the DC-OC transitions of the
multicomponent CLAH models. Analogously, the critical behaviors along
the CH line of the nCLAH models and the DC-OD line of the CLAH models
for any $Q\ge 2$ (these transitions do not exist for $Q=1$) share the
same universality class associated with the stable fixed point of the
AH field theory. The deconfinement transitions we have studied here,
as well as the OC-OD transitions in multicomponent CLAH models for
$Q\ge 4$, belong to the same $XY_G$ universality class as the
topological transitions in nCLAH models (MH line for $N\ge 2$ and CH
line for $N=1$).  Only in CLAH systems with charge $Q=2$ these
deconfinement transitions belong to a different (Ising-gauge)
universality class.

The topological transitions in nCLAH models are associated with the
condensation of a nonlocal order parameter defined by dressing the
scalar field with a nonlocal electromagnetic cloud~\cite{KK-85, KK-86,
  BN-87, BPV-23b, BPV-24}.  The critical correlations of this nonlocal
charged operator were numerically studied in
Refs~\cite{BPV-23b,BPV-24}, exploiting the fact that it assumes a
simplified local form in an appropriate Lorenz gauge, where it reduces
itself to the scalar-field operator. Along the $XY_G$ transition line
they display a power-law critical behavior, which definitely differs
from that of the local scalar order parameter in standard spin $XY$
transitions, thus providing a better characterization of the $XY_G$
universality class.  In these studies the use of the Lorenz gauge
fixing~\cite{BPV-23ybq,BPV-23b} was crucial, as the charged order
parameter becomes local with this choice of gauge.  Since the
topological transitions in CLAH models with $Q\ge 4$ have the same
critical behavior, it is natural to expect that also these transitions
are characterized by the condensation of an appropriately dressed
scalar operator or, equivalently, by a local scalar operator in some
generalized Lorenz gauge.  Such a program is made difficult by the
nonlinear nature of the compact gauge fields that does not allow the
use of simple linear gauge fixings.  A priori one might think of using
other gauge fixings, but they may not be appropriate to investigate
charged nonlocal excitations; see Refs.~\cite{BPV-23ybq,BPV-23b} for a
discussion. For instance, in the unitary gauge no scalar condensation
occurs~\cite{FMS-81}. If a strategy could be found to overcome these
difficulties, it would be possible to investigate charged excitations
both for $Q\ge 4$ (it would represent a further universality check)
and for $Q=2$.

\appendix

\section{Relation between the $\mathbb{Z}_4$ and $\mathbb{Z}_2$
  gauge models}
\label{z4z22}

In this appendix we prove the relation (\ref{zkqrel}) between the
partition functions of the $\mathbb{Z}_4$ and $\mathbb{Z}_2$
gauge models defined by the Hamiltonian (\ref{HZQ}).  For this
purpose, it is convenient to expand the partition function
as follows
\begin{eqnarray}
Z_{\mathbb{Z}_Q}(K)=\sum_{\{\rho\}}e^{-H_{{\mathbb Z}_Q}(K)}=
  \sum_{\{\rho\}}
  \prod_{\mathrm{plaq}}e^{2K \mathrm{Re}\,\Pi_{{\bm x},\mu\nu}},\quad
\end{eqnarray}
where 
  $\Pi_{{\bm x},\mu\nu} = \rho_{{\bm
      x},{\mu}} \,\rho_{{\bm x}+\hat{\mu},{\nu}}
  \,\bar{\rho}_{{\bm x}+\hat{\nu},{\mu}} \,\bar{\rho}_{{\bm  
      x},{\nu}}$.
For $Q=4$ the different values of $\exp(2K \mathrm{Re}\,\Pi_{{\bm
      x},\mu\nu})$, counted with their degeneracies (corresponding to
random choices of $\pm 1$ and $\pm i$ on each link), are
\begin{equation}\label{eq:degQ4}
(Q=4)\quad e^{2K \mathrm{Re}\,\Pi_{{\bm x},\mu\nu}}=\left\{
\begin{array}{ll} e^{2 K} & \mathrm{deg}=64 \\
1 & \mathrm{deg}=128 \\
e^{-2 K} & \mathrm{deg}=64 \end{array}\right. ,
\end{equation}
while for $Q=2$ we have
\begin{equation}
(Q=2)\quad e^{2 K \mathrm{Re}\,\Pi_{{\bm x},\mu\nu}}=\left\{
\begin{array}{ll} e^{2 K} & \mathrm{deg}=8 \\
e^{-2 K} & \mathrm{deg}=8 \end{array}\right. \ .
\end{equation}
We can now write $Z_{\mathbb{Z}_2}(K/2)^2$ as a sum on two independent
sets of configurations, and it is easily seen that the possibile
values (and the corresponding degeneracies) of the expression $\exp[K
  (\mathrm{Re}\,\Pi^{(1)}_{{\bm
      x},\mu\nu}+\mathrm{Re}\,\Pi^{(2)}_{{\bm x},\mu\nu})]$
are exactly the same as reported in Eq.~\eqref{eq:degQ4}, from which
we obtain Eq.~\eqref{zkqrel}.  We mention that an analogous relation
holds for the $\mathbb{Z}_4$ and $\mathbb{Z}_2$ clock
models~\cite{HS-03}.

\section{The role of the boundary conditions in the limiting models}
\label{App.B}

We wish now to discuss the limits $J\to\infty$ and $K\to\infty$ in a
finite volume $L^3$. We consider both periodic boundary conditions
(PBC),
\begin{equation}
\lambda_{{\bm x} + L\hat{\mu},\nu} = \lambda_{{\bm x},\nu} \qquad
z_{{\bm x} + L\hat{\mu}} = z_{\bm x}, 
\end{equation}
and $C^*$ boundary conditions 
\begin{equation}
\lambda_{{\bm x} + L\hat{\mu},\nu} = \overline{\lambda}_{{\bm x},\nu}
\qquad z_{{\bm x} + L\hat{\mu}} = \bar z_{\bm x}.
\end{equation}
The limit $J\to\infty$ is particularly simple. Indeed, if we start
from the unitary-gauge Hamiltonian (\ref{Hunitary}), the increase of
$J$ has the only effect of reducing the fluctuations of $\lambda_{{\bm
    x},\mu}$, which, in the limit, takes only $Q$ different
values. Therefore, the limiting model has the same boundary conditions
considered in the finite-$J$ system.

Let us now discuss the limit $K\to\infty$.  In this limit all gauge
plaquettes are equal to the identity.  For both PBC and $C^*$ boundary
conditions, we can set, modulo gauge transformations,
\begin{eqnarray}
\lambda_{(x_1,x_2,x_3),\mu} &=& V_1 \quad 
    \hbox{if $x_1=L$, $1\le x_2,x_3\le L$, $\mu=1$,} \nonumber \\
    &=& V_2 \quad 
    \hbox{if $x_2=L$, $1\le x_1,x_3\le L$, $\mu=2$,} \nonumber \\
    &=& V_3 \quad 
    \hbox{if $x_3=L$, $1\le x_1,x_2\le L$, $\mu=3$,} \nonumber \\
    &=& 1 \quad \hbox{otherwise},
\label{lambda-Pi1}
\end{eqnarray}
where $V_1,V_2,V_3$ are three arbitrary phases.  We consider here a
cubic lattice of size $L^3$ and set ${\bm x} = (x_1,x_2,x_3)$ with
$1\le x_\mu \le L$.

In the PBC case, the gauge fields given in Eq.~(\ref{lambda-Pi1})
satisfy $\Pi_{{\bm x},\mu\nu} = 1$ on all plaquettes and gauge
invariance is completely fixed. The presence of the three phases
$V_\mu$ is related to the fact that, in the PBC case, a field
configuration $\lambda_{{\bm x},\mu}$ is completely specified (modulo
gauge transformations) not only by the values of the fields on the
plaquettes but also by the values of three Polyakov loops (the product
of the gauge fields along a straight lattice line that winds around
the lattice) in the three different directions. We substitute
Eq.~(\ref{lambda-Pi1}) in the scalar Hamiltonian and obtain the
standard $XY$ vector Hamiltonian except on the boundary links. The
boundary terms are
\begin{eqnarray}
&& -2 J\ \sum_{x_a,x_b}  \hbox{Re} 
  \Bigl( \bar{z}_{(L,x_a,x_b)} z_{(1,x_a,x_b)} W_1  + \\
&& \qquad + \bar{z}_{(x_a,L,x_b)} z_{(x_a,1,x_b)} W_2 +  
   \bar{z}_{(x_a,x_b,L)} z_{(x_a,x_b,1)} W_3\Bigr),
   \nonumber 
\label{boundaryterms}
\end{eqnarray}
where $W_\mu = V_{\mu}^Q$.  Thus, we obtain an $XY$ spin model with
three additional dynamical boundary fields.  They are irrelevant in
the infinite-volume limit, but change the FSS properties of the model
for finite sizes.  One can also view the resulting model as a standard
$XY$ model with fluctuating boundary conditions, i.e., with fields
satisfying
\begin{equation}
z_{{\bm x}+L\hat{\mu}} = W_\mu z_{\bm x}.
\end{equation}
In the case of $C^*$ boundary conditions, Eq.~(\ref{lambda-Pi1}) does
not guarantee that $\Pi_{{\bm x},\mu\nu} = 1$ on all
plaquettes. Indeed, there are three additional conditions.  If we
consider ${\bm x} = (L,L,1)$, $\mu = 1$ and $\nu = 2$, we have
\begin{equation}
\Pi_{{\bm x},\mu\nu} = 
  \lambda_{(L,L,1),1} \bar{\lambda}_{(1,L,1),2} \lambda_{(L,1,1),1} 
  \bar{\lambda}_{(L,L,1),2} = V_1^2 \bar{V}_2^2,
\end{equation}
which implies $V_1^2 = V_2^2$, Analogously, we obtain $V_1^2 = V_3^2$.
Therefore, we can set $V_1 = \sigma_1 V_3$ and $V_2 = \sigma_2 V_3$,
where $\sigma_1$ and $\sigma_2$ can only take the values $\pm 1$.
Substituting the gauge fields in the scalar Hamiltonian, we obtain an
$XY$ model with nontrivial boundary terms:
\begin{eqnarray}
&& -2J\ \sum_{x_a,x_b}  \hbox{Re} 
  \Bigl(z_{L,x_a,x_b} z_{1,x_a,x_b} \sigma_1^Q W \\
&& \qquad  + z_{x_a,L,x_b} z_{x_a,1,x_b} \sigma_2^Q W  + 
  z_{x_a,x_b,L} z_{x_a,x_b,1} W\Bigr), 
\nonumber
\end{eqnarray}
where $W = V_3^Q$. If we now perform a global change of variables
$z'_{\bm x} = W^{1/2} z_{\bm x}$, we can get rid of the phase $W$.  If
$Q$ is even, also the signs $\sigma_1$ and $\sigma_2$ drop out and
thus we obtain an $XY$ spin model with $C^*$ boundary conditions. For
odd $Q$, instead, we obtain an $XY$ model with ${\mathbb
  Z}_2$-fluctuating $C^*$ boundary conditions.

These considerations can also be applied to the AH model with
${\mathbb Z}_2$ gauge and matter fields: $\lambda_{{\bm x},\mu}$ and
$z_{\bm x}$ are real and take only the values $\pm 1$. If we use PBC,
for $J\to\infty$ we obtain the ${\mathbb Z}_2$ lattice gauge model
with the same boundary conditions. For $K\to \infty$ we obtain the
usual Ising model, but with fluctuating boundary conditions.  The
${\mathbb Z}_2$ scalar fields satisfy $z_{{\bm x}+L\hat{\mu}} = V_\mu
{z}_{\bm x}$, where $V_\mu=\pm 1$.  The latter result allows us to
clarify the role of duality in finite-size systems. Ref.~\cite{BDI-75}
showed that the ${\mathbb Z}_2$ AH model satisfies an exact duality
relation for finite $J$ and $K$. The duality relation also holds in a
finite cubic volume if PBC are used. By taking the limits $J\to\infty$
or $K\to\infty$, it is easy to see that duality relates the ${\mathbb
  Z}_2$ pure gauge model obtained for $J\to\infty$ with the Ising
model obtained for $K\to\infty$. This relation also holds in a finite
volume, provided one considers the boundary conditions that are
obtained by properly performing the two limits.  Thus, in a finite
cubic volume the ${\mathbb Z}_2$ pure gauge model with PBC is not
exactly dual to the Ising spin model with PBC. Exact duality holds
only if one considers the Ising model with fluctuating boundary
conditions as discussed above.

\section{Energy-cumulant scaling functions in the 3D ${\mathbb Z}_2$ gauge
model}
\label{fssz2}

\begin{figure}[btp]
  \includegraphics*[width=0.9\columnwidth]{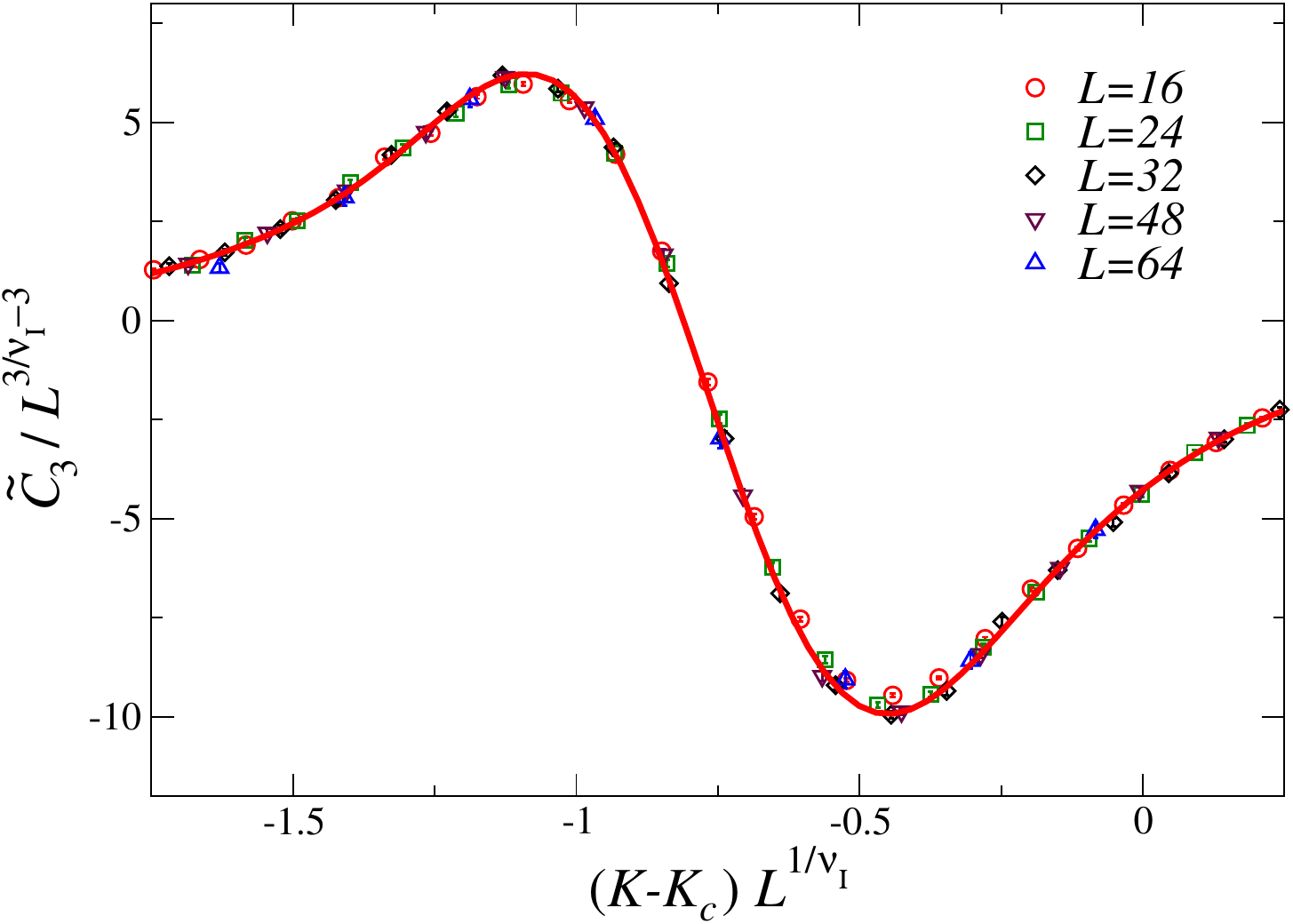}
  \includegraphics*[width=0.9\columnwidth]{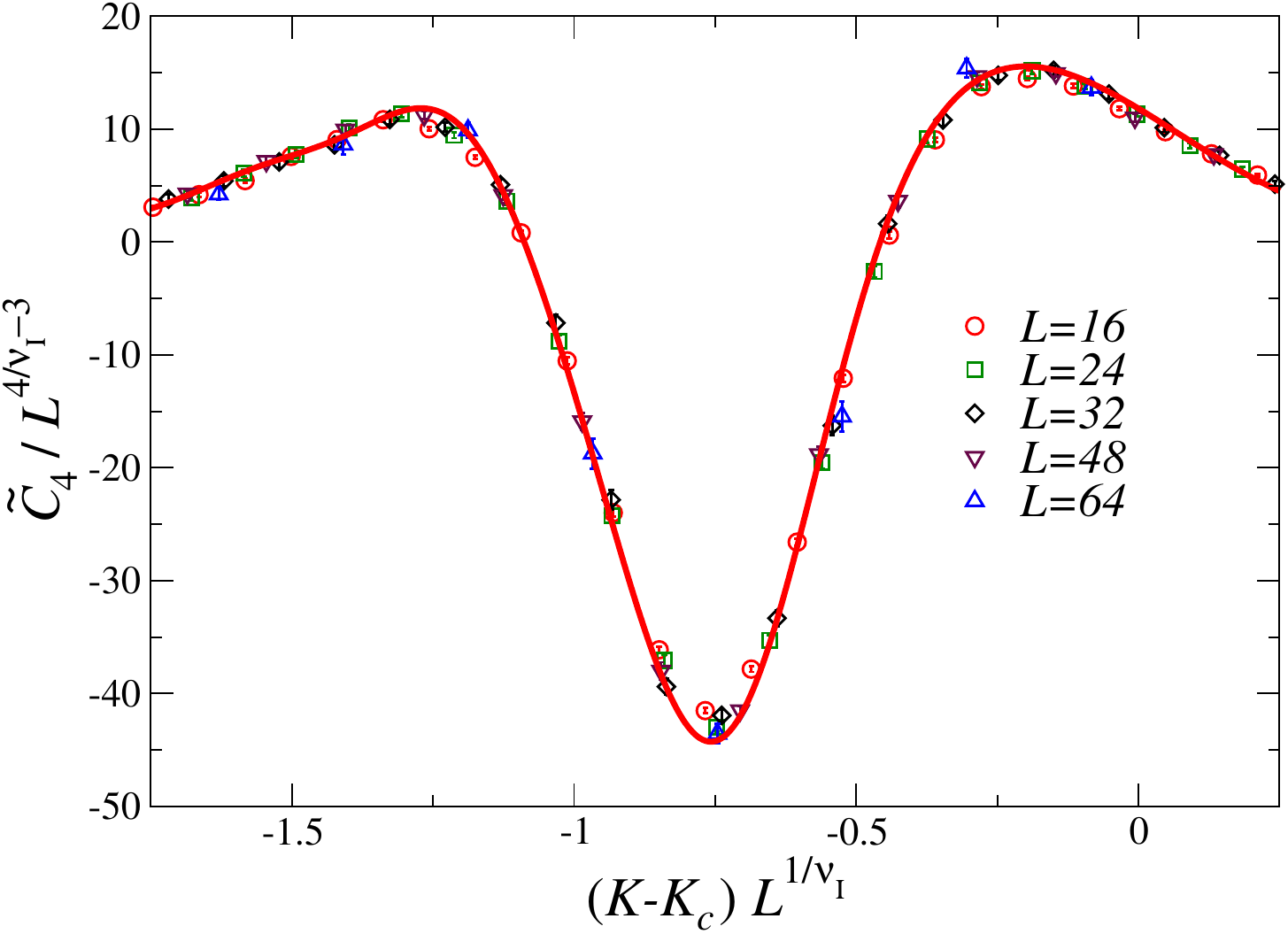}
  \caption{FSS behavior of the energy cumulants $\widetilde{C}_3$
    (top) and $\widetilde{C}_4$ (bottom), for the 3D ${\mathbb Z}_2$
    gauge model.  The solid line in the upper panel is the
    interpolation $\mathcal{U}_3^{(I)}(X)$ reported in
    Eq.~(\ref{interp}), while the solid line in the lower panel
    corresponds to
    $\frac{\mathrm{d}}{\mathrm{d}X}\mathcal{U}_3^{(I)}(X)$.}
\label{z2gauge}
\end{figure}

In this section we determine the universal scaling functions
$\mathcal{U}_k^{(I)}(X)$ for the gauge Ising universality class with
periodic boundary conditions ($C^*$ and periodic boundary conditions
are equivalent for ${\mathbb Z}_2$ gauge variables).  For this purpose
we consider the ${\mathbb Z}_2$ gauge theory and use
Eq.~(\ref{UB-Jinf}).  We have performed high-precision simulations on
lattice up to $L=64$ in the range $-2.0 \le X \le 0.5$.  We have
determined the cumulants $\widetilde{C}_n(K)$ of $H_\lambda$, which
are related to the cumulant $C_n(K)$ defined in Sec.~\ref{fssenecum}
by
\begin{equation}
  \widetilde{C}_n(K,L) = {1\over K^n} C_n(K).
  \label{ctilden}
\end{equation}
Because of Eq.~(\ref{UB-Jinf}) we have in the scaling limit
\begin{equation}
\widetilde{C}_n(K,L) \approx  
   L^{n/\nu - 3} {\mathcal U}_n^{(I)} (X) + \widetilde{b}_n(K).
\end{equation}
The data for $n=3$ and $4$ scale nicely and allow us to compute the
corresponding scaling functions $\mathcal{U}_n^{(I)}(X)$, where we
have added a superscript ${(I)}$ to specify that the results refer to
the Ising gauge universality class with periodic boundary conditions.
Data are quite precisely interpolated by
the following expression:
\begin{widetext}
\begin{eqnarray}
\mathcal{U}_3^{(I)}(X) &=& 
- ( 16399.705 + 
    30410.750 X + 
    26897.615 X^2 + 
    13094.665 X^3 \nonumber \\ 
&&  \quad
     + 3416.983 X^4 + 
      369.113 X^5)
  \exp[-(X - m_1)^2/(2 \sigma_1^2)]
\nonumber \\
&& + (302.042 
- 1329.328 X + 
  2523.811 X^2 - 
  2655.760 X^3 
\nonumber \\    
&&  \quad
  + 1536.383 X^4 - 402.120 X^5) 
  \exp[-(X - m_2)^2/(2 \sigma_2^2)]
\label{interp}
\end{eqnarray}
\end{widetext}
where $m_1 = -1.1$, $m_2 = -0.3$, $\sigma_1 = 0.38$, and $\sigma_2 =
0.46$.  This expression interpolates all data points with $L\ge 32$
with deviations that are of the order of or smaller than the
statistical errors (the errors on $\widetilde{C}_3(K,L) L^{3 -
  3/\nu_I}$ are approximately equal to 0.1 for $L=32,48$ and 0.2 for
$L=64$), see the upper panel of Fig.~\ref{z2gauge}.  By
differentiating $\mathcal{U}_3^{(I)}(X)$ with respect to $X$ we
obtain the scaling curve $\mathcal{U}_4^{(I)}(X)$, which is in very
good agreement with the numerical data, see the lower panel of
Fig.~\ref{z2gauge}. We can compute the universal RG invariant ratios (\ref{Wdef}),
obtaining
\begin{equation}
W_1 = -1.59(3) \qquad
W_2 = 0.41(4) \qquad
W_3 = 0.74(3). 
\end{equation}
These results hold for any model in the Ising gauge universality class with
$C^*$/periodic boundary conditions, and can be compared, e.g., with those
characteristic of the $XY_G$ universality class, see Eq.~\eqref{eq:WIXY}.

\begin{figure}[t]
  \includegraphics*[width=0.9\columnwidth]{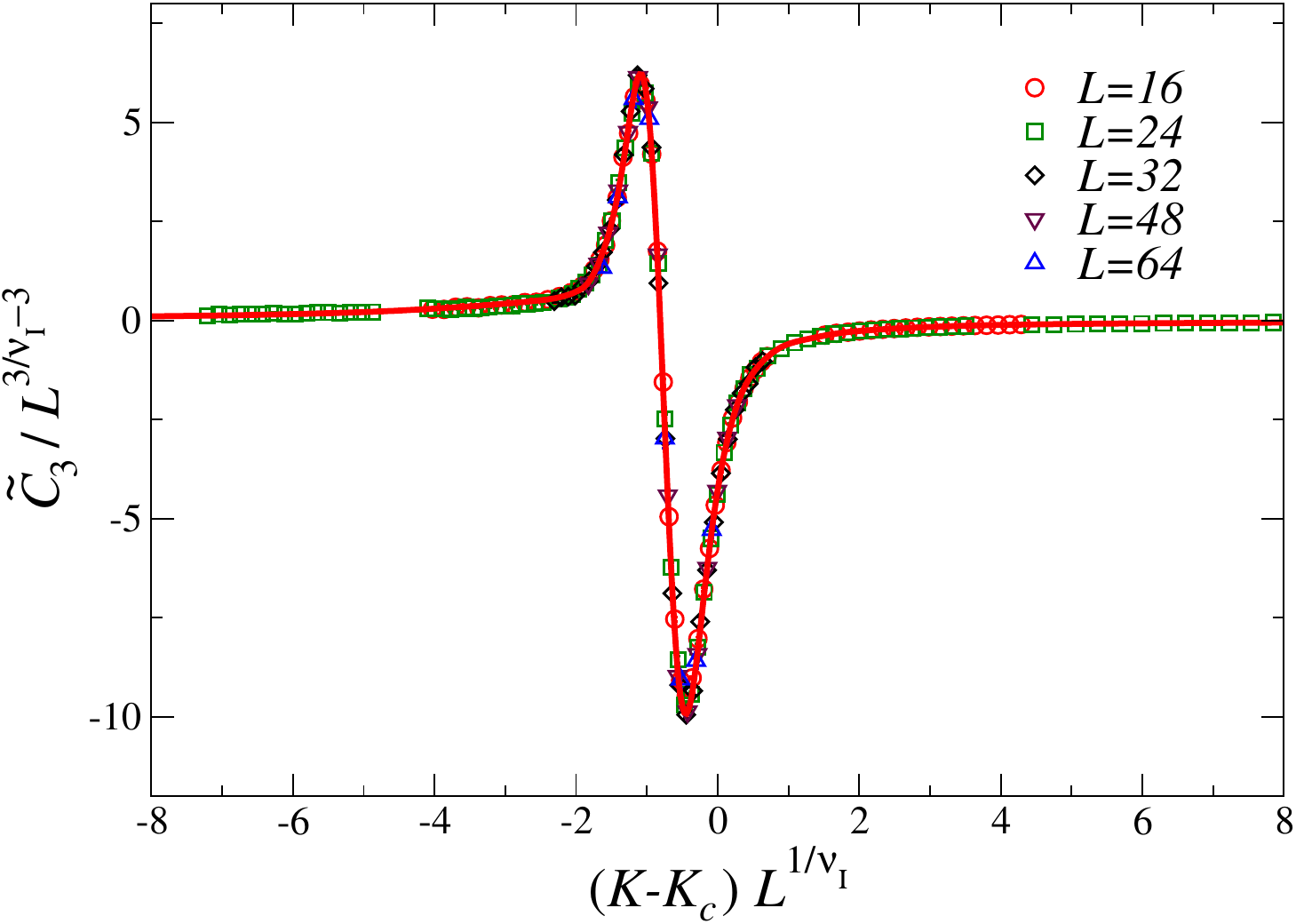}
  \caption{Plot of $\widetilde{B}_3(J,L) L^{-3/\nu_I+3}$ versus
    $X$ in the extended range $-8\le X\le 9$ and interpolated scaling
    curve $\mathcal{U}_3^{(I)}(X)$}
\label{z2gauge-1}
\end{figure}

\begin{figure}[t]
  \includegraphics*[width=0.9\columnwidth]{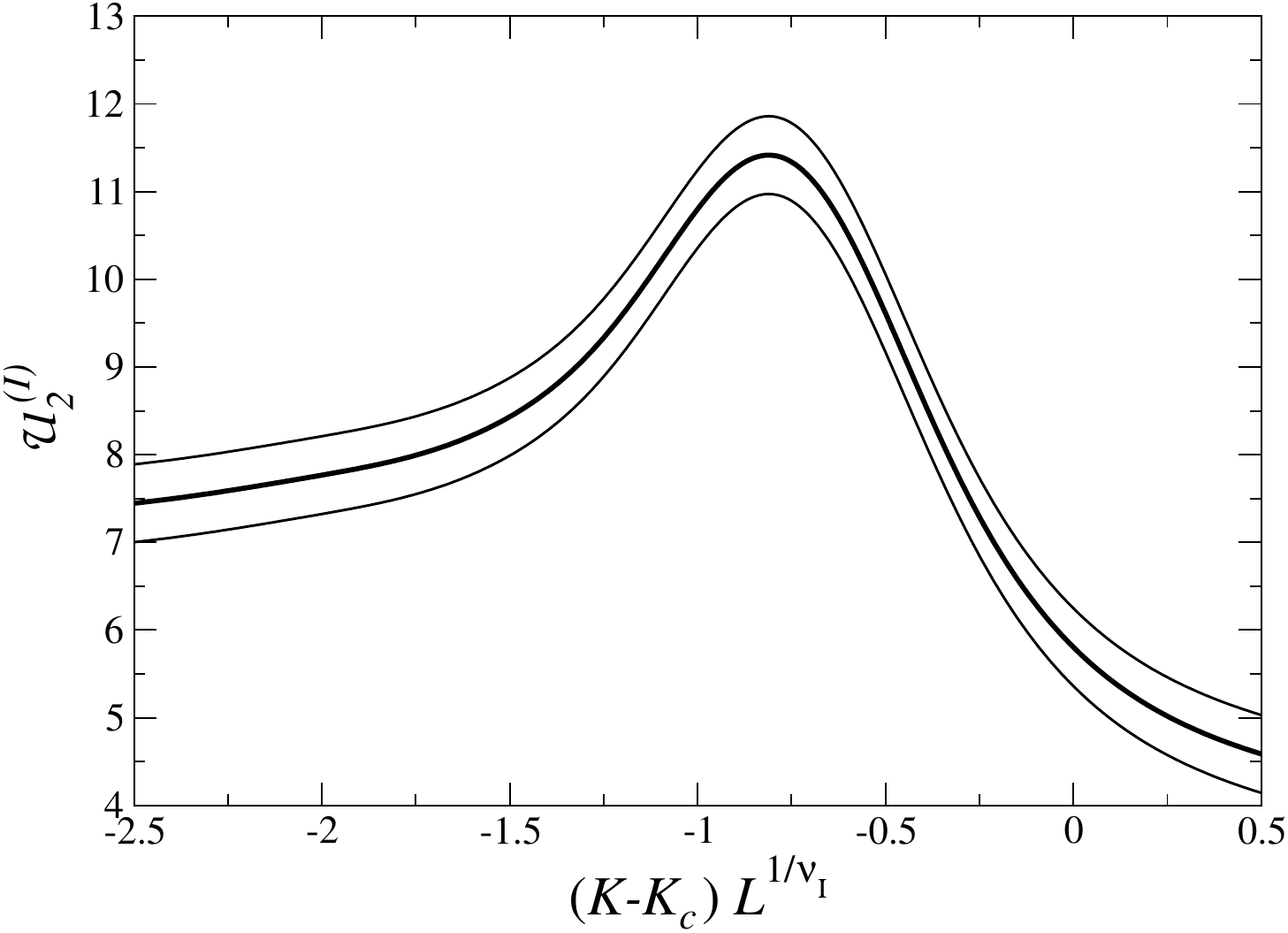}
  \includegraphics*[width=0.9\columnwidth]{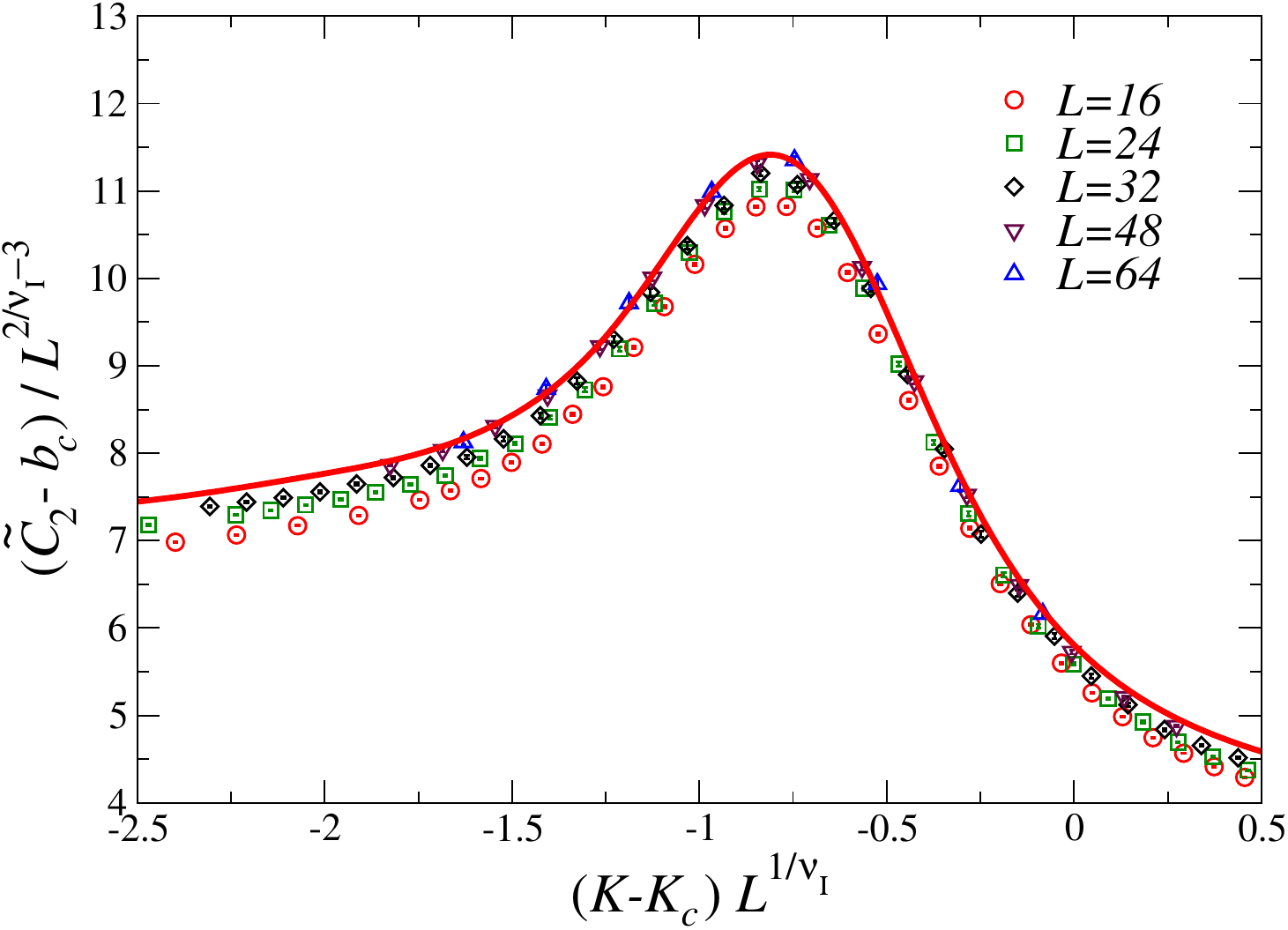}
  \caption{(Top): Plot of the scaling
    curves $\mathcal{U}_{2a}^{(I)}(X)$ and $\mathcal{U}_{2b}^{(I)}(X)$
    (thin lines) and of their average, which is what we consider our
    best estimate (thick central line).  (Bottom): Plot of
    $(\widetilde{C}_2(J,L) - b_c) L^{-2/\nu_I+3}$ with $b_c = -4.10$.
    The thick line is the estimate of $\mathcal{U}_{2}^{(I)}(X)$.  }
\label{z2gauge-2}
\end{figure}

We have also determined the scaling curve $\mathcal{U}_2^{(I)}(X)$ for
the second moment, using relation (\ref{bkint}). For this purpose,
however, we need to determine $\mathcal{U}_3^{(I)}(X)$ on the whole
real axis. In practice, we need simulations in a wider range of values
of $X$, such that we are able to observe the asymptotic behavior
(\ref{largeXBk}) for large values of $|X|$. Then, if we use a
parametrization that satisfies Eq.~(\ref{largeXBk}), we expect it to
be approximately valid also for large and small values of $X$ outside
the simulation interval.  We have therefore performed an additional
set of simulations on lattices of size $L=24$ in the ranges $-8\le X
\le -2$ and $0.6\le X \le 9$ and determined a new parametrization of
the MC data in the whole range $-8\le X \le 9$, see
Fig.~\ref{z2gauge-1}. Then, we have computed two different
approximations of $\mathcal{U}_2^{(I)}(X)$:
\begin{eqnarray}
\mathcal{U}_{2a}^{(I)}(X) &=& \int_{-\infty}^X dY \mathcal{U}_3^{(I)}(Y) 
\nonumber \\
\mathcal{U}_{2b}^{(I)}(X) &=& -\int_X^\infty dY \mathcal{U}_3^{(I)}(Y).
\end{eqnarray}
The difference between the two results allows us to understand the
role of the tails of $\mathcal{U}_3^{(I)}(X)$. Indeed, for values of
$X$ in the simulation interval, the first expression depends on the
extrapolation for $X\to-\infty$, but not on the values of
$\mathcal{U}_3^{(I)}(X)$ for large $X$. Viceversa, the second
expression is only sensitive to the large-$X$ extrapolation. The two
expressions differ by a constant value, which is approximately equal
to 0.9. As our final estimate we consider, see the upper panel of
Fig.~\ref{z2gauge-2},
\begin{equation} 
\mathcal{U}_{2}^{(I)}(X) = {1\over 2}[ \mathcal{U}_{2a}^{(I)}(X) + 
\mathcal{U}_{2b}^{(I)}(X) ].
\end{equation}
We expect the absolute error on this quantity to be approximately
$0.9/2\approx 0.5$. In the interval $-2\le X \le 0.5$, the curve
$\mathcal{U}_{2}^{(I)}(X)$ can be determined using the parametrization
(\ref{interp}) and
\begin{equation}
\mathcal{U}_{2}^{(I)}(X) = \mathcal{U}_{2}^{(I)}(0) + 
     \int_0^X dY \mathcal{U}_{3}^{(I)}(Y),
\end{equation}
where $\mathcal{U}_{2}^{(I)}(0) = 5.8(4)$.

To compare the estimated $\mathcal{U}_{2}^{(I)}(X)$ with the data, it
is necessary to take into account the analytic-background
contribution, see Eq.~(\ref{Hg3-scaling}).  In the scaling limit we
can simply replace $b_2(J)$ with its value at criticality, i.e., with
$b_c = b_2(J_c)/K_c^2$.  This constant has been fixed by requiring the
MC estimates of $\widetilde{C}_2(K,L)$ for $X\approx 0$ and $L\ge 48$
to agree with $\mathcal{U}_{2}^{(I)}(0) L^{2/\nu_I-3} + b_c$.  We
obtain $b_c = -4.1(9)$, where the error is essentially due to the
uncertainty on the scaling curve. Note that background contribution is
large and comparable with the scaling contribution
[$\mathcal{U}_{2}^{(I)}(0) L^{2/\nu_I-3} \approx 11.2$ with an
  approximate error of 0.9, for $L=48$], which explains why it is not
possible to estimate critical exponents from the behavior of $C_2$
(the specific heat). In the lower panel of Fig.~\ref{z2gauge-2}, we
compare $\mathcal{U}_{2}^{(I)}(X)$ with the data, observing a
reasonable agreement for all values of $X$.

\section{Energy-cumulant scaling functions in the inverted $XY$ model}
\label{appXY}

To determine the universal scaling curves ${\cal U}_k^{(XY)}(X)$ for
the $XY_G$ universality class, we should consider an arbitrary model
in this universality class and cumulants normalized so that
Eq.~(\ref{bkrelation}) holds. For this purpose we have considered the
inverted $XY$ model~\cite{DH-81,NRR-03}, which can be obtained as the
$J\to\infty$ limit of the $N=1$ noncompact LAH model, with Hamiltonian
$H = K H_A + J H_z$, where
\begin{eqnarray}
H_A &=& \frac{1}{2} \sum_{{\bm x},\mu>\nu} F_{{\bm x},\mu\nu}^2, 
\nonumber \\
H_z &=& - 2 \sum_{{\bm x},\mu} {\rm Re}\,( \lambda_{{\bm x},\mu} 
  z_{\bm x} z_{{\bm x}+\hat\mu}), 
\label{nCLAHham}
\end{eqnarray}
and partition function
\begin{equation}
Z  = \int [dA_{{\bm x},\mu} d\bar{z}_{\bm x} d{z}_{\bm x}]
e^{-H({\bm A},{z})}.
\end{equation}
We have defined $\lambda_{{\bm x},\mu} = e^{iA_{{\bm x},\mu}}$,
$F_{{\bm x},\mu\nu}= \Delta_{\mu} A_{{\bm x},\nu} - \Delta_{\nu}
A_{{\bm x},\mu}$, $\Delta_\mu A_{{\bm x},\nu} = A_{{\bm
    x}+\hat{\mu},\nu}- A_{{\bm x},\nu}$, and $A_{{\bm x},\nu}$ is a
real variable. This model is well defined on a finite lattice with
$C^*$ boundary conditions~\cite{BPV-21-ncAH}.

\begin{figure}[tbp]
\includegraphics*[width=0.95\columnwidth]{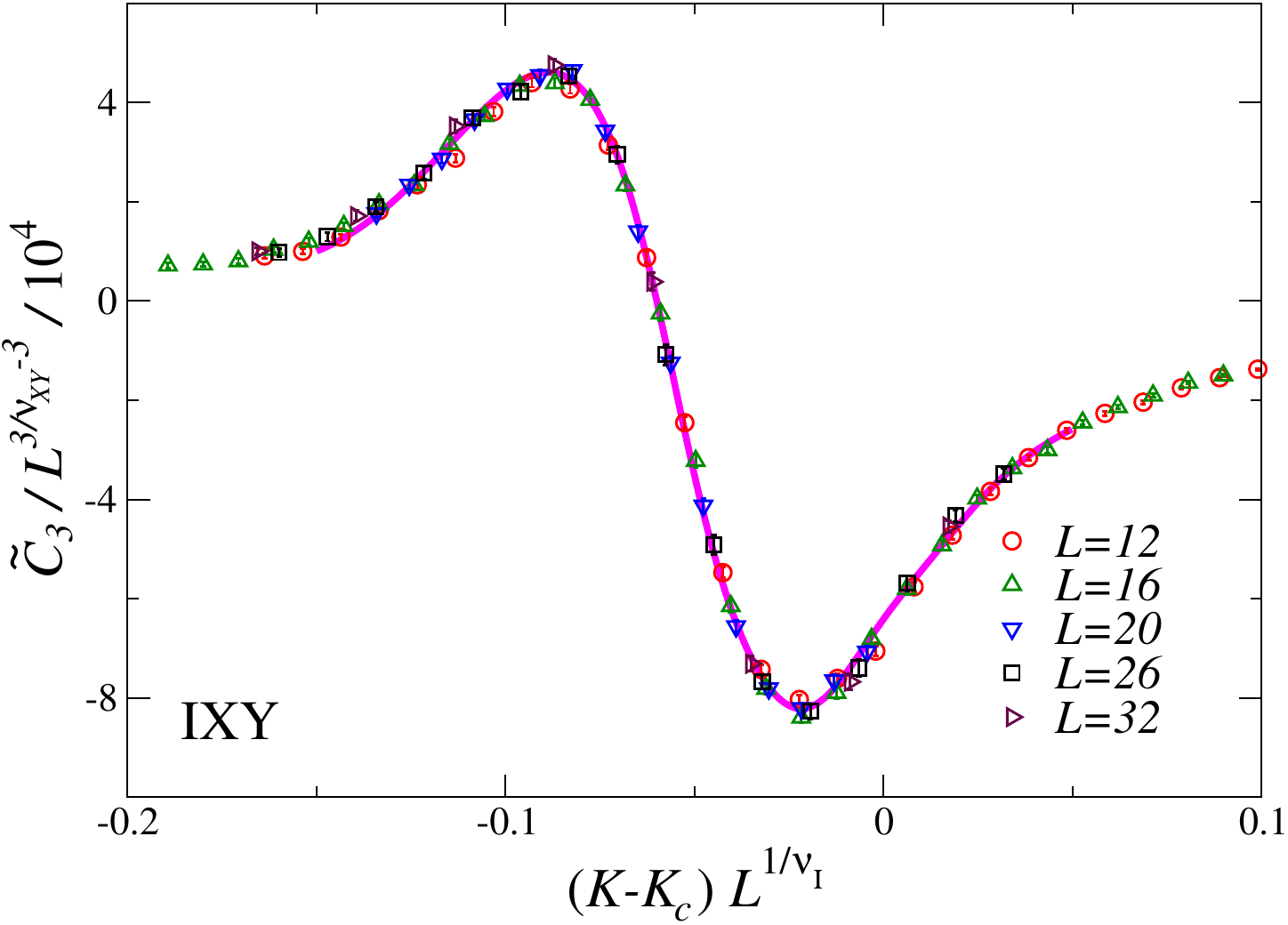}
\caption{Scaling plot of $\widetilde{C}_3$ as a function of $X$ for
  the inverted $XY$ model. We also report the interpolating curve given
  in Eq.~\eqref{eq:B3IXY}.}
\label{fig:B3_scal_IXY}
\end{figure}

We perform simulations in the inverted $XY$ model obtained for large
values of $J$, determining the third cumulant $\widetilde{C}_3$ of
$H_A$, cf. Eq.~(\ref{ctilden}), for lattices up to $L=32$. Data should
scale as
\begin{equation}
\widetilde{C}_3(K,L) \approx  
   L^{3/\nu_{XY} - 3} {\mathcal U}_n^{(XY)} (X) , 
\end{equation}
where $X=(K-K_c)L^{1/\nu_{XY}}$ with $K_c = 0.076051$ (from
Ref.~\cite{NRR-03}). The scaling is excellent, see
Fig.~\ref{fig:B3_scal_IXY}.  The results are parametrized by the
following function:
\begin{widetext}
\begin{equation}\label{eq:B3IXY}
\begin{aligned}
K_c^3 {\cal U}_3^{(XY)}(x)=&
 (104.93856 - 417.6510 x + 2172.749x^2 + 
        72345.33x^3 + 1.4355919\cdot 10^6x^4)
       e^{-0.5 s_2 (x-\mu)^2} \\
&-\frac{150.76378 + 393.6980x + 11444.009x^2}{
        1 + s_2(x-\mu)^2}
+\frac{598978.55x^3 + 9.81037366\cdot 10^6\,x^4}{1 + s_4(x-\mu)^4},
\end{aligned}
\end{equation}
\end{widetext}
where $\mu = -0.021$, $\sigma = 0.042$, $s_2=1/\sigma^2$, and $s_4 =
1/\sigma^4$.  This expression holds for $x$ in the range
$[-0.15,0.05]$. It interpolates quite precisely all data, with an
absolute error $\lesssim 0.5$, The interpolating curve is also
reported in Fig.~\ref{fig:B3_scal_IXY}.

Using these data we can also compute the universal RG invariant ratios 
defined in Eq.~(\ref{Wdef}): 
\begin{equation}
\label{eq:WIXY}
W_1 = -1.75(4) \qquad
W_2 = 0.25(5), \qquad
W_3 = 0.68(5). 
\end{equation}
These results can be compared with the analogous results for the $XY$
spin model with the same boundary conditions. The analysis of the data
with $L=24$ and 32 gives
\begin{equation}
\label{eq:WIXYspin}
W_1 = -2.4(1) \qquad
W_2 = -0.2(1), \qquad
W_3 =  0.5(1). 
\end{equation}

\acknowledgments

The authors acknowledge support from project PRIN 2022 ``Emerging
gauge theories: critical properties and quantum dynamics''
(20227JZKWP).  Numerical simulations have been performed on the CSN4
cluster of the Scientific Computing Center at INFN-PISA.


\begin{thebibliography}{99}

\bibitem{Anderson-book} P. W. Anderson, {\em Basic Notions of
  Condensed Matter Physics}, (The Benjamin/Cummings Publishing
  Company, Menlo Park, California, 1984).

\bibitem{Wen-book} X.-G. Wen, {\em Quantum field theory of many-body
  systems: from the origin of sound to an origin of light and
  electrons}, (Oxford University Press, 2004).

\bibitem{Sachdev-19} S. Sachdev, Topological order, emergent gauge
  fields, and Fermi surface reconstruction, Rep. Prog. Phys. {\bf 82},
  014001 (2019).

\bibitem{HLM-74} B. I. Halperin, T. C. Lubensky, and S. K. Ma,
  First-Order Phase Transitions in Superconductors and Smectic-A
  Liquid Crystals, Phys. Rev. Lett. {\bf 32}, 292 (1974).

\bibitem{FS-79} E. Fradkin and S. Shenker, Phase diagrams of lattice
  gauge theories with Higgs fields, Phys. Rev. D {\bf 19}, 3682
  (1979).

\bibitem{DH-81} C. Dasgupta and B. I. Halperin, Phase Transitions in a 
  Lattice Model of Superconductivity,
  Phys. Rev. Lett {\bf 47}, 1556 (1981).

\bibitem{FMS-81} J. Fr\"ohlich, G. Morchio, and F. Strocchi, Higgs
  phenomenon without symmetry breaking order parameter,
  Nucl. Phys. {\bf B190}, 553 (1981).

\bibitem{DHMNP-81} P. Di Vecchia, A. Holtkamp, R. Musto, F. Nicodemi,
  and R. Pettorino, Lattice CP$^{N-1}$ models and their large-$N$
  behaviour, Nucl. Phys. B {\bf 190}, 719 (1981).

\bibitem{BF-81} G. Bhanot and B. A. Freedman, Finite size scaling for
  the 3D abelian Higgs model, Nucl. Phys. B {\bf 190}, 357 (1981).

\bibitem{CC-82} D. J. E. Callaway and L. J. Carson, Abelian Higgs
  model: A Monte Carlo study, Phys. Rev. D {\bf 25}, 531 (1982).

\bibitem{BF-83} J. Bricmont and J. Fr\"ohlich, An order parameter
  distinguishing between different phases of lattice gauge theories
  with matter fields, Phys. Lett. {\bf 122}, 73 (1983).

\bibitem{FM-83} K. Fredenhagen and M. Marcu, Charged states in $Z_2$
  gauge theories, Commun. Math. Phys. {\bf 92}, 81 (1983).

\bibitem{KK-85} T. Kennedy and C. King, Symmetry Breaking in the
  Lattice Abelian Higgs Model, Phys. Rev. Lett. {\bf 55}, 776 (1985).

\bibitem{KK-86} T. Kennedy and C. King, Spontaneous Symmetry Breakdown
  in the Abelian Higgs Model, Commun. Math. Phys. {\bf 104}, 327
  (1986).

\bibitem{BN-86} C. Borgs and F. Nill, Symmetry Breaking in Landau
  Gauge A comment to a paper by T. Kennedy and C. King,
  Commun. Math. Phys. {\bf 104}, 349 (1986).

\bibitem{BN-86-b} C. Borgs and F. Nill, No Higgs mechanism in scalar
  lattice QED with strong electromagnetic coupling, Phys. Lett. B {\bf
    171}, 289 (1986).

\bibitem{BN-87} C. Borgs and F. Nill, The Phase Diagram of the Abelian
  Lattice Higgs Model. A Review of Rigorous Results,
  J. Stat. Phys. {\bf 47}, 877 (1987).

\bibitem{RS-90} N. Read and S. Sachdev, Spin-Peierls, valence-bond
  solid, and N\'eel ground states of low-dimensional quantum
  antiferromagnets, Phys. Rev. B {\bf 42}, 4568 (1990).

\bibitem{MS-90} G. Murthy and S. Sachdev, Actions of hedgehogs
  instantons in the disordered phase of 2+1 dimensional CP$^{N-1}$
  model, Nucl. Phys. B {\bf 344}, 557 (1990).

\bibitem{KKS-94} M. Kiometzis, H. Kleinert, and A. M. J. Schakel,
  Critical Exponents of the Superconducting Phase Transition,
  Phys. Rev. Lett.  {\bf 73}, 1975 (1994).

\bibitem{BFLLW-96} B. Bergerhoff, F. Freire, D.F. Litim, S. Lola, and
  C. Wetterich, Phase diagram of superconductors from nonperturbative
  flow equations, Phys. Rev. B {\bf 53}, 5734 (1996).
  
\bibitem{HT-96} F. Herbut and Z. Tesanovic, Critical Fluctuations in
  Superconductors and the Magnetic Field Penetration Depth,
  Phys. Rev. Lett. {\bf 76}, 4588 (1996).

\bibitem{FH-96} R. Folk and Y. Holovatch, On the critical fluctuations
  in superconductors, J. Phys. A {\bf 29}, 3409 (1996).

\bibitem{IKK-96} V. Yu. Irkhin, A. A. Katanin, and M. I. Katsnelson,
  $1/N$ expansion for critical exponents of magnetic phase transitions
  in the $CP^{N-1}$ model for $2<d<4$, Phys. Rev. B {\bf 54}, 11953
  (1996).
  
\bibitem{KKLP-98} K. Kajantie, M. Karjalainen, M. Laine, and J. Peisa,
  Masses and phase structure in the Ginzburg-Landau model,
  Phys. Rev. B {\bf 57}, 3011 (1998).

\bibitem{OT-98} P. Olsson and S. Teitel, Critical Behavior of the
  Meissner Transition in the Lattice London Superconductor,
  Phys. Rev. Lett. {\bf 80}, 1964 (1998).

\bibitem{CN-99} C. de Calan and F.S. Nogueira, Scaling critical
  behavior of superconductors at zero magnetic field, Phys. Rev. B
  {\bf 60}, 4255 (1999)
  
\bibitem{HS-00} J. Hove and A. Sudbo, Anomalous Scaling Dimensions and
  Stable Charged Fixed Point of Type-II Superconductors,
  Phys. Rev. Lett. {\bf 84}, 3426 (2000).

\bibitem{SSS-02} R. D.  Sedgewick, D. J. Scalapino, and R. L. Sugar,
  Fractionalized phase in an $XY-Z_2$ gauge model, Phys. Rev. B {\bf
    65}, 054508 (2002).

\bibitem{SM-02} T. Senthil and O. Motrunich, Microscopic models for
  fractionalized phases in strongly correlated systems, Phys. Rev. B
  {\bf 66}, 205104 (2002).

\bibitem{KNS-02} H. Kleinert, F. S. Nogueira, and A. Sudb{\o},
  Deconfinement Transition in Three-Dimensional Compact U(1) Gauge
  Theories Coupled to Matter Fields, Phys. Rev. Lett. {\bf 88}, 232001
  (2002).

\bibitem{MHS-02} S. Mo, J. Hove, and A. Sudb\o, Order of the
  metal-to-superconductor transition, Phys. Rev. B {\bf 65}, 104501
  (2002).

\bibitem{SSSNH-02} A. Sudb{\o}, E. Sm{\o}rgrav, J. Smiseth,
  F. S. Nogueira, and J. Hove, Criticality in the (2+1)-Dimensional
  Compact Higgs Model and Fractionalized Insulators,
  Phys. Rev. Lett. {\bf 89}, 226403 (2002).

\bibitem{SSNHS-03} J. Smiseth, E. Sm{\o}rgrav, F. S. Nogueira,
  J. Hove, and A. Sudb{\o}, Phase Structure of $d = 2 + 1$ Compact
  Lattice Gauge Theories and the Transition from Mott Insulator to
  Fractionalized Insulator, Phys. Rev. B {\bf 67}, 205104 (2003).

\bibitem{MZ-03} M. Moshe and J. Zinn-Justin, Quantum field theory in
  the large $N$ limit: A review, Phys. Rep. {\bf 385}, 69 (2003).

\bibitem{NRR-03} T. Neuhaus, A. Rajantie, and K. Rummukainen,
  Numerical study of duality and universality in a frozen
  superconductor, Phys. Rev. B {\bf 67}, 014525 (2003).

\bibitem{SBSVF-04} T. Senthil, L. Balents, S. Sachdev, A. Vishwanath,
  and M. P. A. Fisher, Quantum Criticality beyond the
  Landau-Ginzburg-Wilson Paradigm, Phys. Rev. B {\bf 70}, 144407
  (2004).
  
\bibitem{MV-04} O. I. Motrunich and A. Vishwanath, Emergent photons
  and transitions in the O(3) $\sigma$-model with hedgehog
  suppression, Phys. Rev. B {\bf 70}, 075104 (2004).

\bibitem{NSSS-04} F. S. Nogueira, J. Smiseth, E. Sm{\o}rgrav, and
  A. Sudb{\o}, Compact U(1) gauge theories in $2+1$ dimensions and the
  physics of low dimensional insulating materials, Eur. Phys. J. C
  {\bf 33}, 885 (2004).

\bibitem{SSS-04} J. Smiseth, E. Smorgrav, and A. Sudb{\o}, Critical
  properties of the $N$-color London model, Phys. Rev. Lett. {\bf 93},
  077002 (2004).

\bibitem{TIM-05} S. Takashima, I. Ichinose, and T. Matsui, CP$^1$+U(1)
  lattice gauge theory in three dimensions: Phase structure, spins,
  gauge bosons, and instantons, Phys. Rev. B {\bf 72}, 075112 (2005).

\bibitem{TIM-06} S. Takashima, I. Ichinose, and T. Matsui,
  Deconfinement of spinons on critical points: Multiflavor CP$^1$+U(1)
  lattice gauge theory in three dimension, Phys. Rev. B {\bf 73},
  075119 (2006).
  
\bibitem{WBJSS-05} S. Wenzel, E. Bittner, W.  Janke, A. M. J. Schakel,
  and A. Schiller, Kertesz Line in the Three-Dimensional Compact U(1)
  Lattice Higgs Model, Phys. Rev. Lett. {\bf 95}, 051601 (2005).

\bibitem{CFIS-05} M. N. Chernodub, R. Feldmann, E.-M. Ilgenfritz, and
  A. Schiller, The compact $Q = 2$ Abelian Higgs model in the London
  limit: vortex-monopole chains and the photon propagator,
  Phys. Rev. D {\bf 71}, 074502 (2005).

\bibitem{HW-05} M. B. Hastings and X.-G. Wen, Quasi-adiabatic
  Continuation of Quantum States: The Stability of Topological Ground
  State Degeneracy and Emergent Gauge Invariance,
  Phys. Rev. B {\bf 72}, 045141 (2005)
  
\bibitem{CIS-06} M. N. Chernodub, E.-M. Ilgenfritz, and A. Schiller,
  Phase structure of an Abelian two-Higgs model and high temperature
  superconductors, Phys. Rev. B {\bf 73}, 100506 (2006).

\bibitem{KPST-06} A. B. Kuklov, N. V. Prokof'ev, B. V. Svistunov, and
  M. Troyer, Deconfined criticality, runaway flow in the two-component
  scalar electrodynamics and weak first-order superfluid-solid
  transitions, Ann. Phys. \textbf{321}, 1602 (2006).

\bibitem{Sandvik-07} A. W. Sandvik, Evidence for Deconfined Quantum
  Criticality in a Two-Dimensional Heisenberg Model with Four-Spin
  Interactions, Phys. Rev. Lett. {\bf 98}, 227202 (2007).

\bibitem{Herbut-book} I. Herbut, \textit{A Modern Approach to Critical
  Phenomena} (Cambridge University Press, 2007).

\bibitem{MK-08} R. G. Melko and R. K. Kaul, 
  Scaling in the Fan of an Unconventional Quantum Critical Point, 
  Phys. Rev. Lett. {\bf 100}, 017203 (2008).

\bibitem{JNCW-08} F. J. Jiang, M. Nyfeler, S. Chandrasekharan, and
  U. J. Wiese, From an Antiferromagnet to a Valence Bond Solid:
  Evidence for a First-Order Phase Transition, J. Stat. Mech. (2008)
  P02009.

\bibitem{WBJS-08} S. Wenzel, E. Bittner, W. Janke, and
  A. M. J. Schakel, Percolation of Vortices in the 3D Abelian Lattice
  Higgs Model, Nucl. Phys. B {\bf 793}, 344 (2008).
  
\bibitem{MV-08} O. I. Motrunich and A. Vishwanath, Comparative study
  of Higgs transition in one-component and two-component lattice
  superconductor models, arXiv:0805.1494 [cond-mat.stat-mech].

\bibitem{KMPST-08} A. B. Kuklov, M. Matsumoto, N. V. Prokof'ev,
  B. V. Svistunov, and M. Troyer, Deconfined Criticality: Generic
  First-Order Transition in the SU(2) Symmetry Case,
  Phys. Rev. Lett. {\bf 101}, 050405 (2008).

\bibitem{CAP-08} D. Charrier, F. Alet, and P. Pujol, Gauge Theory
  Picture of an Ordering Transition in a Dimer Model,
  Phys. Rev. Lett. {\bf 101}, 167205 (2008).
  
\bibitem{KS-08} R. K. Kaul and S. Sachdev, Quantum criticality of U(1)
  gauge theories with fermionic and bosonic matter in two spatial
  dimensions, Phys. Rev. B {\bf 77}, 155105 (2008).

\bibitem{ODHIM-09} T. Ono, S. Doi, Y. Hori, I. Ichinose, and
  T. Matsui, Phase Structure and Critical Behavior of Multi-Higgs U(1)
  Lattice Gauge Theory in Three Dimensions, Ann. Phys. (N.Y.) {\bf
    324}, 2453 (2009).

\bibitem{LSK-09} J. Lou, A. W. Sandvik, and N. Kawashima,
  Antiferromagnetic to valence-bond-solid transitions in
  two-dimensional SU(N) Heisenberg models with multispin interactions,
  Phys. Rev. B {\bf 80}, 180414 (2009).

\bibitem{CGTAB-09} G. Chen, J. Gukelberger, S. Trebst, F. Alet, and
  L. Balents, Coulomb gas transitions in three-dimensional classical
  dimer models, Phys. Rev. B {\bf 80}, 045112 (2009).

\bibitem{Sandvik-10} A. W. Sandvik, Continuous Quantum Phase
  Transition between an Antiferromagnet and a Valence-Bond Solid in
  Two Dimensions: Evidence for Logarithmic Corrections to Scaling,
  Phys. Rev. Lett. {\bf 104}, 177201 (2010).

\bibitem{CA-10} D. Charrier and F. Alet, Phase diagram of an extended
  classical dimer model, Phys. Rev. B {\bf 82}, 014429 (2010).

\bibitem{BDA-10} A. Banerjee, K. Damle, and F. Alet, Impurity spin
  texture at a deconfined quantum critical point, Phys. Rev. B {\bf
    82}, 155139 (2010).

\bibitem{GS-10}
  T. Grover and T. Senthil,
  Quantum phase transition from an antiferromagnet to a spin liquid
  in a metal, Phys. Rev. B {\bf 81}, 205102 (2010).

\bibitem{IMH-12} S. V. Isakov, R. G. Melko, and M. B. Hastings,
  Universal signatures of fractionalized quantum critical points,
  Science {\bf 335}, 193 (2012).

\bibitem{Kaul-12} R. K. Kaul, Quantum phase transitions in bilayer
  SU($N$) antiferromagnets, Phys. Rev. B {\bf 85}, 180411(R) (2012).

\bibitem{KS-12} R. K. Kaul and A. W. Sandvik, Lattice Model for the
  SU($N$) N\'eel to Valence-Bond Solid Quantum Phase Transition at
  Large $N$, Phys. Rev. Lett. {\bf 108}, 137201 (2012).

\bibitem{HSOMLWTK-13} K. Harada, T. Suzuki, T. Okubo, H. Matsuo,
  J. Lou, H. Watanabe, S. Todo, and N. Kawashima, Possibility of
  Deconfined Criticality in SU($N$) Heisenberg Models at Small $N$,
  Phys. Rev. B {\bf 88}, 220408 (2013).

\bibitem{CHDKPS-13} K. Chen, Y. Huang, Y. Deng, A. B. Kuklov,
  N. V. Prokof'ev, and B.V. Svistunov, Deconfined Criticality Flow in
  the Heisenberg Model with Ring-Exchange Interactions,
  Phys. Rev. Lett. {\bf 110}, 185701 (2013).

\bibitem{PDA-13} S. Pujari, K. Damle, and F. Alet, N\`eel-State to
  Valence-Bond-Solid Transition on the Honeycomb Lattice: Evidence for
  Deconfined Criticality, Phys. Rev. Lett. {\bf 111}, 087203 (2013).

\bibitem{BMK-13} M. S. Block, R. G. Melko, and R. K. Kaul, Fate of
  CP$^{N-1}$ fixed point with $q$ monopoles, Phys. Rev. Lett. {\bf
    111}, 137202 (2013).
  
\bibitem{HBBS-13} E. V. Herland, T. A. Bojesen, E. Babaev, and
  A. Sudb\o, Phase structure and phase transitions in a
  three-dimensional SU(2) superconductor, Phys. Rev. B {\bf 87},
  134503 (2013).

\bibitem{Bartosch-13} L. Bartosch, Corrections to scaling in the
  critical theory of deconfined criticality, Phys. Rev. B {\bf 88},
  195140 (2013).

\bibitem{BS-13} T. A. Bojesen and A. Sudb\o, Berry phases, current
  lattices, and suppression of phase transitions in a lattice gauge
  theory of quantum antiferromagnets, Phys. Rev. B {\bf 88}, 094412
  (2013).

\bibitem{NSCOS-15} A. Nahum, P. Serna, J. T. Chalker, M. Ortu\`no, and
  A. M. Somoza, Emergent SO(5) Symmetry at the N\'eel to
  Valence-Bond-Solid Transition, Phys. Rev. Lett. {\bf 115}, 267203
  (2015).

\bibitem{NCSOS-15} A. Nahum, J. T. Chalker, P. Serna, M. Ortu\`no, and
  A. M. Somoza, Deconfined Quantum Criticality, Scaling Violations,
  and Classical Loop Models, Phys. Rev. X {\bf 5}, 041048 (2015).

\bibitem{SP-15} G. J. Sreejith and S. Powell, Scaling dimensions of
  higher-charge monopoles at deconfined critical points, Phys. Rev. B
  {\bf 92}, 184413 (2015).

\bibitem{SGS-16} H. Shao, W. Guo, and A. W. Sandvik, Quantum
  criticality with two length scales, Science {\bf 352}, 213 (2016).
  
\bibitem{WNMXS-17} C. Wang, A. Nahum, M. A. Metliski, C. Xu, and
  T. Senthil, Deconfined Quantum Critical Points: Symmetries and
  Dualities, Phys. Rev. X {\bf 7}, 031051 (2017).
  
\bibitem{FH-17} G. Fejos and T. Hatsuda, 
  Renormalization group flows of the $N$-component Abelian Higgs model, 
  Phys. Rev. D {\bf 96}, 056018 (2017).

\bibitem{IZMHS-19} B. Ihrig, N. Zerf, P. Marquard, I. F. Herbut, and
  M. M. Scherer, Abelian Higgs model at four loops, fixed-point
  collision and deconfined criticality, Phys. Rev. B {\bf 100}, 134507
  (2019).

\bibitem{SN-19} P. Serna and A. Nahum, Emergence and spontaneous
  breaking of approximate O(4) symmetry at a weakly first-order
  deconfined phase transition, Phys. Rev. B {\bf 99}, 195110 (2019).

\bibitem{PV-19-CP} A. Pelissetto and E. Vicari, 
  Three-dimensional ferromagnetic CP$^{N-1}$ models, 
  Phys. Rev. E {\bf 100}, 022122 (2019).

\bibitem{PV-19-AH3d} A. Pelissetto and E. Vicari, 
  Multicomponent compact Abelian-Higgs lattice models, 
  Phys. Rev. E {\bf 100}, 042134 (2019).
  
\bibitem{SZ-20} A. W. Sandvik and B. Zhao, Consistent scaling
  exponents at the deconfined quantum-critical point,
  Chin. Phys. Lett. {\bf 37}, 057502 (2020).

\bibitem{PV-20-mfcp} A. Pelissetto and E. Vicari, 
  Three-dimensional monopole-free CP$^{N-1}$ models, 
  Phys. Rev. E {\bf 101}, 062136 (2020).

\bibitem{PV-20-largeNCP} A. Pelissetto and E. Vicari, 
  Large-$N$ behavior of three-dimensional lattice CP$^{N-1}$ models, 
  J. Stat. Mech.: Th. Expt.  033209 (2020).

\bibitem{BPV-20-hcAH} C. Bonati, A. Pelissetto, and E. Vicari,
  Higher-charge three-dimensional compact lattice Abelian-Higgs models, 
  Phys. Rev. E {\bf 102}, 062151 (2020).

\bibitem{BPV-21-ncAH} C.~Bonati, A.~Pelissetto, and E.~Vicari, Lattice
  Abelian-Higgs model with noncompact gauge fields, Phys. Rev. B {\bf
    103}, 085104 (2021).

\bibitem{BPV-21} C. Bonati, A. Pelissetto and E. Vicari, Breaking of
  Gauge Symmetry in Lattice Gauge Theories, Phys. Rev. Lett. {\bf
    127}, 091601 (2021)

\bibitem{BPV-21-bgs} C. Bonati, A. Pelissetto, and E. Vicari, Lattice
  gauge theories in the presence of a linear gauge-symmetry breaking,
  Phys. Rev. E {\bf 104}, 014140 (2021).

\bibitem{WB-21}
  D. Weston and E. Babaev,
  Composite order in SU(N) theories coupled to an Abelian gauge field,
  Phys. Rev. B {\bf 104}, 075116 (2021).
  
\bibitem{BPV-22-mpf} C. Bonati, A. Pelissetto, and E. Vicari,
  Three-dimensional monopole-free CP$^{N-1}$ models: Behavior in the
  presence of a quartic potential, J. Stat. Mech. (2022) 063206.

\bibitem{BPV-22} C. Bonati, A. Pelissetto, and E. Vicari, Critical
  behaviors of lattice U(1) gauge models and three-dimensional
  Abelian-Higgs gauge field theory, Phys. Rev. B {\bf 105}, 085112
  (2022).

\bibitem{BPV-22-dis} C. Bonati, A. Pelissetto, and E. Vicari, Scalar
  gauge-Higgs models with discrete Abelian symmetry group,
  Phys. Rev. E {\bf 105} 054132 (2022).
  
\bibitem{BPV-23b} C. Bonati, A. Pelissetto, and E. Vicari,
  Coulomb-Higgs phase transition of three-dimensional lattice Abelian
  Higgs gauge models with noncompact gauge variables and gauge fixing
  Phys. Rev. E \textbf{108}, 044125 (2023).
  
\bibitem{SZJSM-23}
  M. Song, J. Zhao, L. Janssen, M. M. Scherer, and
  Z. Y. Meng, Deconfined quantum criticality lost,
  arXiv:2307.02547.

\bibitem{BPV-23d} C. Bonati, A. Pelissetto, and E. Vicari, Abelian
  Higgs gauge theories with multicomponent scalar fields and
  multiparameter scalar potentials, Phys. Rev. B {\bf 108}, 245154
  (2023).

\bibitem{BPV-24} C. Bonati, A. Pelissetto, and E. Vicari, Diverse
  universality classes of the topological deconfinement transitions of
  three-dimensional noncompact lattice Abelian-Higgs models,
  Phys. Rev. D (2024) in press, arXiv:2308.00101.

\bibitem{Baxter-book}  
 R. J. Baxter, {\em Exactly Solved Model in Statistical Mechanics},
  Academic Press, London, 1982.
  
\bibitem{AT-43} J. Ashkin and E. Teller, Statistics of two-Dimensional
  lattices with four components, Phys. Rev. {\bf 64}, 178 (1943).

\bibitem{KW-71} L. P. Kadanoff and F. J. Wagner, Some critical
  properties of the eight-vertex model, Phys. Rev. B. 4, 3989 (1971).

\bibitem{MM-23} I. Mukherjee and P. K. Mohanty, Hidden super
  universality in systems with continuous variation of critical
  exponents, arXiv:2307.15022.

\bibitem{KT-73} 
J. M. Kosterlitz and  D. J. Thouless,
  Ordering, metastability and phase transitions in two-dimensional systems,
  J.\ Phys. C: Solid State {\bf 6},  1181 (1973).

\bibitem{Berezinskii-70} V. L. Berezinskii, Destruction of Long-range
  Order in One-dimensional and Two-dimensional Systems having a
  Continuous Symmetry Group I. Classical Systems,
  Zh. Eksp. Theor. Fiz. {\bf 59}, 907 (1970) [Sov. Phys. JETP {\bf
      32}, 493 (1971)].

\bibitem{Kosterlitz-74}
  J. M. Kosterlitz, 
  The critical properties of the two- dimensional xy model,
  J. Phys. C {\bf 7}, 1046 (1974).

\bibitem{JKKN-77} J. V. Jos\'e, L. P. Kadanoff, S. Kirkpatrick, and
  D. R. Nelson, Renormalization, vortices, and symmetry-breaking
  perturbations in the two-dimensional planar model, Phys. Rev. B
  {\bf 16}, 1217 (1977).

\bibitem{Wegner-71} F. J. Wegner, Duality in Generalized Ising Models
  and Phase Transitions Without Local Order Parameters,
  J. Math. Phys. \textbf{12}, 2259 (1971).

\bibitem{Bhanot:1980pc} G. Bhanot and M. Creutz, The Phase Diagram of
  $Z$(n) and $u$(1) Gauge Theories in Three-dimensions, Phys. Rev. D
  {\bf 21}, 2892 (1980).

\bibitem{Borisenko:2013xna} O. Borisenko, V. Chelnokov, G. Cortese,
  M. Gravina, A. Papa, and I. Surzhikov, Critical behavior of 3D Z(N)
  lattice gauge theories at zero temperature, Nucl. Phys. B
  {\bf 879}, 80 (2014).

\bibitem{Savit-80} R. Savit,
  Duality in Field Theory and Statistical Systems,
  Rev. Mod. Phys. \textbf{52}, 453 (1980).

\bibitem{Aharony-76} A. Aharony, Dependence of universal critical
  behavior on symmetry and range of interaction, in {\em Phase
    transitions and critical phenomena}, vol. 6, p. 357, edited by
  C. Domb and M. S. Green (Academic Press, London, 1976).
\bibitem{PV-02} A. Pelissetto and E. Vicari, 
  Critical Phenomena and Renormalization Group Theory, 
  Phys. Rep. {\bf 368}, 549 (2002).

\bibitem{CPV-00}
J. Carmona, A. Pelissetto, and E. Vicari, The $N$-component
Ginzburg-Landau Hamiltonian with cubic symmetry: a six-loop study,
Phys. Rev. B {\bf 61}, 15136 (2000).

\bibitem{Hasenbusch-23} M. Hasenbusch, Cubic fixed point in three
  dimensions: Monte Carlo simulations of the $\varphi^4$ model on the
  lattice, Phys. Rev. B {\bf 107}, 024409 (2023).
  
\bibitem{CPV-03} P. Calabrese, A. Pelissetto, and E. Vicari,
  Multicritical behavior of $O(n_1)\oplus O(n_2)$-symmetric systems,
  Phys. Rev. B {\bf 67}, 054505 (2003).

\bibitem{HV-11} M. Hasenbusch and E. Vicari, Anisotropic perturbations
  in 3D O(N) vector models, Phys. Rev. B {\bf 84}, 125136 (2011).

\bibitem{KPSV-16} F. Kos, D. Poland, D. Simmons-Duffin, and A. Vichi,
  Precision islands in the Ising and O($N$) models,
  J. High Energy Phys. JHEP 08 (2016) 036.  

\bibitem{HS-03} J. Hove and A. Sudb{\o}, Criticality versus $q$ in the
  (2+1)-dimensional $Z_q$ clock model, Phys. Rev. E {\bf 68}, 046107
  (2003).
  
\bibitem{FXL-18} A. M. Ferrenberg, J. Xu, and D. P. Landau, Pushing
  the limits of Monte Carlo simulations for the three-dimensional
  Ising model, Phys. Rev. E {\bf 97}, 043301 (2018).
  
\bibitem{Hasenbusch-19} M. Hasenbusch, Monte Carlo study of an
  improved clock model in three dimensions, Phys. Rev. B {\bf 100},
  224517 (2019).  

\bibitem{CHPV-06} M. Campostrini, M. Hasenbusch, A. Pelissetto, and
  E. Vicari, Theoretical estimates of the critical exponents of the
  superfluid transition in $^4$He by lattice methods, Phys. Rev. B
  {\bf 74}, 144506 (2006).

\bibitem{DBN-05} Y. Deng, H. W. J. Bl\"ote, and M. P. Nightingale,
  Surface and bulk transitions in three-dimensional $O(N)$ models,
  Phys. Rev. E {\bf 72}, 016128 (2005).
  
\bibitem{BPV-22-z2g} C. Bonati, A. Pelissetto, and E. Vicari,
  Multicritical point of the three-dimensional Z2 gauge Higgs model,
  Phys. Rev. B {\bf 105}, 165138 (2022).

\bibitem{KW-91} A.~S.~Kronfeld and U.~J.~Wiese, SU(N) gauge theories
  with C periodic boundary conditions. 1. Topological structure,
  Nucl. Phys. B {\bf 357}, 521 (1991).

\bibitem{LPRT-16} B.~Lucini, A.~Patella, A.~Ramos, and N.~Tantalo,
  Charged hadrons in local finite-volume QED+QCD with C$^*$ boundary
  conditions, JHEP {\bf 02}, 076 (2016).
  
\bibitem{over1} S. L. Adler, 
  An Overrelaxation Method for the Monte Carlo Evaluation of the Partition
  Function for Multiquadratic Actions,
  Phys. Rev. D \textbf{23}, 2901 (1981)

\bibitem{metro} N. Metropolis, A. W. Rosenbluth,
  M. N. Rosenbluth, A. H. Teller, and E. Teller, 
  Equation of state calculations by fast computing machines, 
  J. Chem. Phys. {\bf 21}, 1087 (1953).  

\bibitem{GZ-98} R. Guida and J. Zinn-Justin, Critical exponents of the
  $N$-vector model, J. Phys. A {\bf 31}, 8103 (1998).

\bibitem{CPRV-02} M. Campostrini, A. Pelissetto, P. Rossi, and
  E. Vicari, 25th order high-temperature expansion results for three-
  dimensional Ising-like systems on the simple cubic lattice,
  Phys. Rev. E {\bf 65}, 066127 (2002).

\bibitem{Hasenbusch-10} M. Hasenbusch, Finite-size scaling study of
  lattice models in the three-dimensional Ising universality class,
  Phys. Rev. B {\bf 82}, 174433 (2010).

\bibitem{KP-17} M. V. Kompaniets and E. Panzer, Minimally subtracted six-loop 
  renormalization of $\phi^4$-symmetric theory and critical exponents, 
  Phys. Rev. D {\bf 96}, 036016 (2017).  

\bibitem{Hasenbusch-21}
  M. Hasenbusch, Restoring isotropy in a three-dimensional lattice model:
  The Ising universality class, Phys. Rev. B {\bf 104}, 014426 (2021).

\bibitem{CLLPSSV-20} S. M. Chester, W. Landry, J. Liu, D. Poland,
  D. Simmons-Duffin, N. Su, and A. Vichi, Carving out OPE space and
  precise O(2) model critical exponents, J. High Energy Phys. {\bf
    06}, 142 (2020).

\bibitem{BPV-23ybq}
  C.~Bonati, A.~Pelissetto and E.~Vicari,
  Gauge fixing and gauge correlations in noncompact Abelian gauge models,
  Phys. Rev. D \textbf{108}, 014517 (2023).

\bibitem{BDI-75} R. Balian, J. M. Drouffe, and C. Itzykson, Gauge
  fields on a lattice. II. Gauge-invariant Ising model, Phys. Rev. D
  {\bf 11}, 2098 (1975). 


\end{thebibliography}
\end{document}